\documentclass[fleqn,usenatbib]{mnras}

\usepackage{newtxtext,newtxmath}

\usepackage[T1]{fontenc}

\DeclareRobustCommand{\VAN}[3]{#2}
\let\VANthebibliography\thebibliography
\def\thebibliography{\DeclareRobustCommand{\VAN}[3]{##3}\VANthebibliography}

\usepackage{graphicx}	
\usepackage{amsmath}	

\usepackage[dvipsnames]{xcolor}
\usepackage{ulem}

\newcommand\akin{\alpha_{\rm k}}
\newcommand\am{$\alpha$-$M$}
\newcommand\amag{\alpha_{\rm m}}
\newcommand\cs{c_{\rm s}}
\newcommand\Ek{E_{\rm k}}
\newcommand\Eg{E_{\rm g}}
\newcommand\Em{E_{\rm m}}

\newcommand\kms{\mbox{\,km s}^{-1}}
\newcommand\Lcal{{\cal{L}}}
\newcommand\Lcalv{{\cal{L}}_v}
\newcommand\Lcalb{{\cal{L}}_{\rm{m}}}
\newcommand\Lcalbmean{{\cal{L}}_{\rm m, avg}}
\newcommand\Lcalrms{{\cal{L}}_{\rm m,rms}}
\newcommand\LS{$\Lcal$-$\Sigma$}
\newcommand\Msun{M_{\odot}}
\newcommand\nth{n_{{\rm th}}}
\newcommand\pcc{\mbox{\,cm}^{-3}}
\newcommand\psc{\mbox{\,cm}^{-2}}
\newcommand\sigmav{\sigma_v}
\newcommand\K{\rm K}
\newcommand{\ppcc}{{\rm cm}^{-3}}
\newcommand{\mG}{\mu {\rm G}}
\newcommand{\Bavg}{B_{\rm avg}}
\newcommand{\Brms}{B_{\rm rms}}

\title[Kinetic and magnetic energy budget in hub-filament systems]{The kinetic and magnetic energy budget of hub-filament systems during the gravitational fragmentation of molecular clouds}

\author[Camacho et al.]{
Vianey Camacho,$^{1,2}$\thanks{E-mail: v.camacho@irya.unam.mx}
Enrique Vázquez-Semadeni,$^{1}$ Aina Palau,$^{1}$ 
and Manuel Zamora-Avilés$^{2}$
\\
$^{1}$Instituto de Radioastronom\'ia y Astrof\'isica, Antigua Carretera a P\'atzcuaro 8701, Ex-Hda. San Jos\'e de la Huerta, 58089 Morelia, Michoac\'an, M\'exico \\
$^{2}$Instituto Nacional de Astrof\'isica, \'Optica y Electr\'onica, Luis E. Erro 1, 72840 Tonantzintla, Puebla, M\'exico\\
}

\date{Accepted XXX. Received YYY; in original form ZZZ}

\pubyear{2021}

\begin{document}
\label{firstpage}
\pagerange{\pageref{firstpage}--\pageref{lastpage}}
\maketitle

\begin{abstract}
We present a numerical study of the balance between the gravitational ($\Eg$), kinetic ($\Ek$), and magnetic ($\Em$) energies of structures within a hub-filament system in a simulation of the formation and global hierarchical collapse (GHC) of a giant molecular cloud. 
For structures defined by various density thresholds, and at different evolutionary stages, we investigate the scaling of the virial parameter, $\alpha$, with mass $M$, and of the {\it Larson ratio}, $\Lcalv \equiv \sigmav/R^{1/2}$, with column density $\Sigma$, where $\sigmav$ is the 1D velocity dispersion, and $R$ is an effective radius.
We also investigate these scalings for the corresponding magnetic parameters $\amag$ and $\Lcalb$.
Finally, we compare our numerical results with an observational sample of massive clumps. 
We find that:
1) $\amag$ and $\Lcalb$ follow similar \am\ and \LS\ scalings as their kinetic counterparts, although the ratio $\Em/\Ek$ decreases as $|\Eg|$ increases. 
2) The largest objects, defined by the lowest thresholds, tend to appear gravitationally bound (and magnetically supercritical), while their internal substructures tend to appear unbound (and subcritical). This suggests that the latter are being compressed by the infall of their parent structures, and supports earlier suggestions that the measured mass-to-magnetic flux ratio $\mu$ decreases inwards in a centrally-peaked cloud under ideal MHD. 
3)~The scatter in the $\alpha$-$M$ and $\Lcal$-$\Sigma$ plots is reduced when $\Ek$ and $\Em$ are plotted directly against $\Eg$, suggesting that the scatter is due to an ambiguity between mass and size. 
4) The clumps in our GHC simulation follow the same trends as the observational sample of massive clumps in the \LS\ and \am\ diagrams. We conclude that 
the main controlling parameter of the energy budget in the structures is $\Eg$, with the kinetic and magnetic energies being derived from it.
\end{abstract}




\section{Introduction}
\label{sec:intro}

\subsection{{\bf Background}} \label{sec:bgd}

The possibility that molecular clouds are in a state of global gravitational collapse \citep{Goldreich+74, Liszt+74} has resurfaced by a series of numerical and observational results \citep[e.g.,][]{Heitsch+09, VS+09, VS+19, Zamora+14, Ibanez+16, Hacar+17, Juarez+17, BP+18, Traficante+18a, Hu+21}.  
In particular, a model that allows  
various mechanisms (e.g., turbulence, accretion, and stellar feedback) to take part of the star formation process, from the largest to the smallest scales, is the gravitational hierarchical collapse scenario, hereafter GHC \citep{VS+19}, in which a state of equilibrium is not assumed and clouds and their substructures (hereinafter generically called clumps) are thought of as non-equilibrium, non-supported entities, undergoing multi-scale gravitational contraction, hierarchically embedded within one another. 

According to this scenario, 
the virial appearance of molecular clouds and their substructure (clumps and cores), which has been interpreted before as an equilibrium state against collapse, is an indicator of the evolutionary state of the clouds. In the GHC model, self-gravity acts at all scales promoting local collapses that occur as consequence of the large-scale gravitational contraction of the parent cloud, where 
Larson's scaling relations \citep{Larson81} constitute a special case of a more general relation between the Larson ratio $\Lcal$ \citep{Keto+86, Heyer+09, BP+11, BP+18, Leroy+15, Camacho+16} and the column density $\Sigma$. The Larson ratio is defined as
\begin{equation}
\Lcal \equiv \frac{\sigmav}{R^{1/2}},
\end{equation}
 where $\sigmav$ is the one-dimensional velocity dispersion and  $R$ is an effective radius, The column density is computed as $\Sigma=M/(\pi R^2)$, where $M$ is the mass of the cloud. 

Assuming the simplified form for a uniform-density sphere, the gravitational energy $\Eg$ is expressed as
\begin{equation}
\Eg = - \frac{3 GM^2}{5R},
\label{eq:Eg}
\end{equation}
while the kinetic energy $\Ek$ for the one-dimensional velocity dispersion is given by
\begin{equation}
\Ek = \frac{3 M \sigma_v^2}{2}.
\label{eq:Ek}
\end{equation}
Now, assuming virial equilibrium between the kinetic and gravitational energies, $|E_\mathrm{g}|=2E_\mathrm{k}$, $\Lcal$ is related to the column density by
\begin{equation}
\Lcalv \equiv \frac{\sigmav}{R^{1/2}} =  \left(\frac{\pi G\Sigma} {5}\right)^{1/2}.
\label{eq:lrat} 
\end{equation} 
Therefore, it is customary to investigate the nearness to virial equilibrium (or, more generally, energy equipartition) by plotting the Larson ratio {\it versus} the column density \citep[e.g.,] [] {Leroy+15, Miville+17, Traficante+18a, Traficante+18b, BP+18}, and we will hereafter refer to this plot as the \LS\ (or \textit{Keto-Heyer}) diagram \citep{Keto+86, Heyer+09}.

A more frequently used indicator of the balance between the gravitational and the kinetic energy is the virial parameter $\alpha$, defined as \citep{Bertoldi+92}
\begin{equation}
\alpha \equiv \frac{2 \Ek} {|\Eg|}
\label{eq:def_alpha}
\end{equation}
and it is customary to plot this parameter against the mass of the clumps and cores \citep[e.g.,] [] {Leroy+15, Miville+17, Traficante+18a, Traficante+18b}. We thus refer to this plot as the $\alpha$-$M$ diagram.
\begin{figure*}
    \includegraphics[width=2\columnwidth]{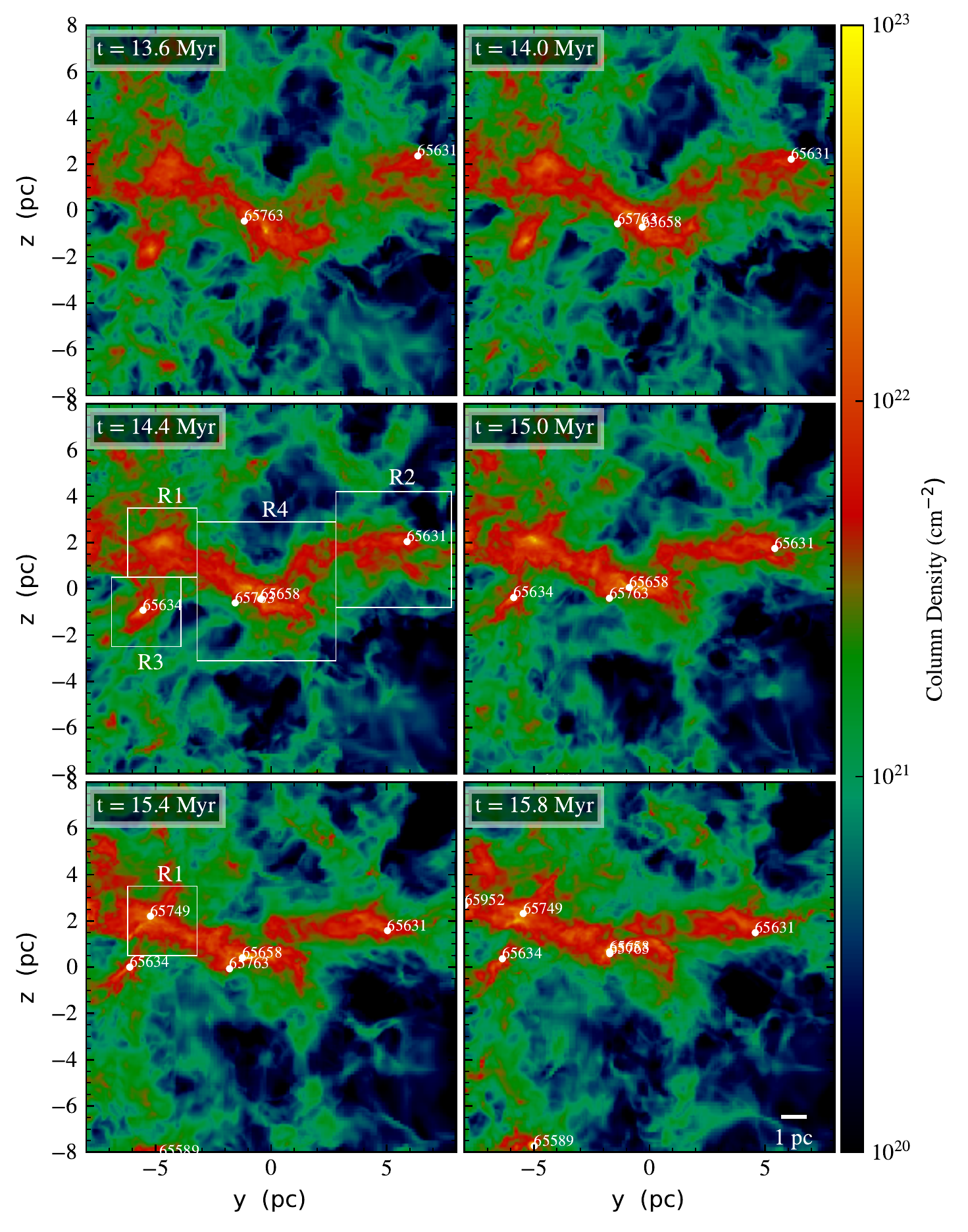}
\caption{Column density maps of the filamentary region at various moments in its evolution, from time $t=13.6$ to $t=15.8$ Myr. The filament is studied at t=14.4 Myr both as a single cloud complex and as a set of four separate subregions with different levels of star formation, labeled R1-R4 according to their star formation activity. The numbers identify the existing sink particles (ID) in each region.}
\label{fig:sim}
\end{figure*}

In previous papers, we have studied the energy balance predicted by the GHC scenario for a set of clumps in numerical simulations where clouds are formed self-consistently due to gas accretion from the diffuse medium to the initial dense turbulent fluctuations, and later driven to collapse by their self-gravity. 
In \citet[][hereafter Paper I]{Camacho+16}, we showed that, even though the simulations were in gravitational contraction,  most clouds with low column density appeared supervirial, although roughly half of those were in the process of assembly, while the rest were in the process of dispersal. Furthermore, in \citet[][hereafter, Paper II]{Camacho+20}, we tracked the evolution of the same clumps studied in Paper I in the $\Lcal-\Sigma$ diagram,  to investigate how the energy generated by gravitational collapse is transferred to the kinetic modes. We showed that this energy evolves, while simultaneously the clumps increase their star formation activity, and found good consistency with observational data of two clouds at different evolutionary stages. In general, we found that the energy budget of clumps and cores is consistent with a process of gravitational collapse from the large scales, triggered by large-scale converging motions that assemble the clumps, causing the clumps' self-gravity to become increasingly dominant as the column density of the clumps increases.
We tested our model against observational data and found good agreement with our results. Given that the simulations were performed to achieve the hierarchical collapse scenario without continuous turbulent driving, we concluded that the GHC model is consistent with observations.
Furthermore, as proposed in the GHC scenario, the clumps in these simulations grow in mass and size during their evolution, causing the dynamics and the star-formation activity to evolve simultaneously.

However, these studies were developed without considering stellar feedback or magnetic fields. To  avoid biases due to the lack of feedback, the  analyses in our previous works were performed at relatively early stages of the clumps' evolution, during which the effects of feedback are not expected to be significant. The magnetic field, on the other hand, was not considered. Thus, an important question is how these results change in the magnetic case.

The magnetic field is generally considered as a potential factor to delay collapse and reduce star formation and this is at the heart of current debates. Observations in general show that diffuse cold gas (atomic and molecular cloud envelopes) tends to be magnetically subcritical (i.e., magnetically supported against self-gravity), while the denser parts of molecular clouds and their substructures tend to be magnetically supercritical \citep[i.e., {\it not} supported by the field;] [] {Crutcher12, Pattle+22}, which has been also observed in numerical studies \citep[e.g.,][]{Ibanez+22}. This is consistent with the diffuse atomic gas accumulating preferentially along magnetic field lines by external driving mechanisms (e.g., the stellar spiral arm gravitational potential, supernova-driven or generic turbulent compressions, etc.) until it becomes magnetically supercritical and molecular, and begins to contract gravitationally \citep{Hartmann+01, VS+11}.
In the present work, we explore how the energy generated by gravitational collapse is distributed among the kinetic and magnetic energies and how it contributes to the clumps' energy budget.

In the kinetic case, the energy balance can be expressed as $a\Ek+\Eg=0$, where $a=2$ for virial equilibrium or $a=1$ in the free-fall case (Papers I, II). 
In the magnetic case, \citet[][]{Mouschovias+95} showed, for a set of clouds analyzed by \citet{Myers88a}, that the Larson ratio squared is proportional to the column density. \citet{Mouschovias+95} interpreted this result as a consequence of the clouds being near magnetic criticality, which implies that the column density and the magnetic field are linearly related.

In this work we study the balance between the gravitational ($\Eg$), kinetic ($\Ek$), and magnetic ($\Em$) energies of clumps in an AMR simulation of a molecular cloud undergoing global hierarchical collapse (GHC) in the $\Lcal-\Sigma$ (or Keto-Heyer) diagram, with the aims of testing whether a simulation in this regime is consistent with observations; of investigating the relative importance of the magnetic field along the cloud-to-core hierarchy in hub-filament systems, and of investigating the origin and the development of the \LS\ and the \am\ scaling relations. In section \ref{sec:method}, we define various quantities and describe the simulation and the clump sample. Section \ref{sec:results} describes the results of the comparison between the different energies. Finally, in sections \ref{sec:disc} and \ref{sec:concl} we present the discussions and conclusions, respectively.

\section{Methodology} 
\label{sec:method}

\subsection{Definitions} \label{sec:defs}

In order to put the magnetic field in comparable terms to the velocity field, here we define the magnetic counterparts of the budget parameters $\Lcal$ and $\alpha$. We start by considering the magnetic energy $\Em$, given by
\begin{equation}
\Em= \frac{1}{8\pi}\int {B}^2 {dV},
\label{eq:Em}
\end{equation}
and the equipartition condition between magnetic and kinetic energies,
\begin{equation}
\Em = |\Eg| \Rightarrow \frac{{B^2}}{18 \pi^2} = \frac{G\Sigma^2}{5}. 
\end{equation}
We therefore define the {\it magnetic Larson ratio}, by analogy to eq.\ \eqref{eq:lrat}, as
\begin{equation}
\Lcal_{m} \equiv \frac{{B}}{(18 \pi \Sigma)^{1/2}},
\label{eq:lmag} 
\end{equation} 
so that the equipartition condition becomes
\begin{equation}
\Lcal_{m} = \left(\frac{\pi G \Sigma} {5} \right)^{1/2}.
\label{eq:mag_equip} 
\end{equation} 

It is important to note that $\Lcalb$ can either be estimated using the magnitude of the {\it mean} magnetic field, 
\begin{equation}
\Bavg =|(1/V)\int_V  \mathrm{B} dV|, 
\end{equation}
or the {\it rms} value, 
\begin{equation}
\Brms = \left[(1/V)\int_V B^2 dV\right]^{1/2},
\end{equation}
which includes the energy of the magnetic fluctuations. Thus, we define $\Lcal_{m, \mathrm{avg}}$ and $\Lcal_{m,\mathrm{rms}}$ respectively. 

Similarly to the Larson ratio, the virial parameter can also be expressed in the kinetic and magnetic case. The kinetic virial parameter is defined as
\begin{equation}
  \akin= \frac{2\Ek}{|\Eg|} \approx \frac{5 \sigmav^2 R} {GM},
  \label{eq:ak}
\end{equation}
and the {\it magnetic} virial parameter as 
\begin{equation}
 \amag \equiv \frac{\Em}{|\Eg|}. 
 \label{eq:am}
\end{equation}
Note that we define $\amag$ {\it without} the factor of 2 that goes into its kinetic counterpart because the magnetic energy appears without this factor in the virial theorem. With this definition, the condition of viral equilibrium corresponds to $\akin = \amag =1$. However, equipartition between the kinetic and gravitational energies implies $\akin =2$, while equipartition between the magnetic and gravitational energies also corresponds to $\amag =1$.

Another point to notice is that $\Em/|\Eg| = 5 B^2 R^4/18\pi^2GM^2$, 
which can be related to the magnetic flux, $\phi \approx \pi R^2 B$, as: 
\begin{equation}
    \amag = \frac{\Em}{|\Eg|} \gtrsim \frac{5}{18 \pi^2 G} \left(\frac{\phi}{M} \right)^2 \equiv \frac{1}  {\mu^2},
    \label{eq:fmr}
\end{equation}
where $\mu$ is the mass-to-magnetic flux ratio normalized to the critical value $\sqrt{5/18 \pi^2 G}$ obtained from the virial theorem
for a uniform sphere. In the relation between the second and third members of the above expression, the equality holds for a uniform magnetic field. Thus, the magnetic virial parameter is bounded from below by the inverse squared mass-to-flux ratio. 

\subsection{Numerical simulation}
\label{sec:simulation}

The numerical simulation used in the present study was performed with the AMR FLASH (v2.5) code \citep{flash} to study the formation and evolution of a magnetized molecular cloud. The initial conditions are as follows: the numerical box, of sizes 
$L_{x}=256$ and $L_{y,z}=128$ pc, is initially filled with a warm neutral gas in thermal equilibrium, with uniform density of $2 \, \ppcc$, and constant temperature of $1450 \, \K$. We set up two cylindrical flows in the $x$-direction, of length 112 pc each, which collide at the center of the numerical simulation at a velocity of 7.5 $\kms$. We add a subsonic turbulent velocity field (with a Mach Number of 0.7) to break the symmetry in the shocked layer and thus trigger dynamical instabilities that generate turbulence within the nascent cloud. 
The gas in the shocked layer suffers a transition to the cold neutral phase due to nonlinear triggering of the thermal instability \citep[e.g.,] [] {Henneb_Perault99, KI00}, and it eventually acquires densities and temperatures corresponding to molecular gas, becoming gravitationally unstable, while continuing to accrete external gas. However, we do not explicitly follow the chemistry in the simulation. 

We dynamically refine according to the Jeans criterion \citep{Truelove+97} up to a maximum resolution of $0.007$ pc, in such a way that we resolve the Jeans length by at least 16 cells. We start checking for sink formation once the local density reaches the threshold number density of $10^7 \, \ppcc$. Once we form a sink, this can accrete mass from their surroundings \citep[][]{Federrath+10b}. 
The numerical box is initially permeated by a uniform magnetic field of $3 \, \mG$ along the $x$-direction.
The total mass-\-to-\-flux ratio in the cylinders is $1.59$ times the critical value, so that the cloud formed by the colliding flows eventually will become mag\-ne\-ti\-cally supercritical once it accretes enough mass. The numerical model used here is identical to that presented in \citet[][model labeled B3J]{Zamora+18}, but with four times higher resolution. For further references and details of the numerical model, we refer the reader to that paper.

For the present work, we select a box of size $L_{\rm box}= 16$ pc and total mass $M \approx 3961 \Msun$ at $t = 14.4$ Myr, containing a large well-defined filament. The filament consists of various regions labeled as R1-R4 according to their star formation level. These regions are indicated by the square boxes at $t=14.4$ Myr panel of Fig.\ \ref{fig:sim}. The most massive sink, located in region R4, has $71.2\,\Msun$ (ID 65763), while the other sinks have $13.3 \,\Msun$ (ID 65658, also in R4), $5.6\,\Msun$ (ID 65634, in R3), and $1.8\,\Msun$ (ID 65631, in R2). Region R1 contains no sinks at this time yet, although it accretes enough mass to form them later $t=15.4$ panel  in Fig.\ \ref{fig:sim}). 
Furthermore, we observe that R1, which starts its star formation activity 1 Myr later than R4, becomes significantly denser and becomes a second hub in the filament. Thus, the complex can be considered as a hierarchical hub-filament system, with R4 and R1 being the hubs, and R2 and R3 being secondary star-forming regions in the filamentary cloud \citep[see, e.g.,] [] {GV14, VS+19}.
Finally, in order to identify the clouds and clumps, in each of the four regions, we interpolate the original non-uniform grid from the AMR simulation into a uniform grid at the highest resolution (0.007 pc).

\subsection{Clump and core sample}
\label{sec:sample}

Figure \ref{fig:sim} shows the column density map, at different times, of the hub-filament system described in the previous section. 
Similarly to Paper I, generic clumps and cores (density enhancements at any density level, and called clumps here for simplicity) are selected in each region as 
connected regions above certain density thresholds.\footnote{We do not use a branching-based algorithm such as {\sc Dendrograms} because we prefer the clumps in the different regions to be defined at consistent density thresholds, rather than depending on when a structure branches off.} We consider threshold densities of $\nth = 300, 10^3, 3\times10^3, 10^4, 3\times10^4,$ and $10^5 \pcc$.

Finally, for each clump, we compute its mass as
$M = V_{\rm cell} \sum \rho$, where $V_{\rm cell}$ is the (uniform) cell volume, $\rho$ is the cell density, and the sum goes over all cells within the clump, its representative ``radius'' $R = (3V/4\pi)^{1/3}$, and its one dimensional velocity dispersion $\sigmav$, as the velocity standard deviation over $\sqrt{3}$.
The kinetic and magnetic Larson ratios $\Lcal_\mathrm{v}$ and $\Lcal_\mathrm{m}$, are defined by equations \eqref{eq:lrat} and \eqref{eq:lmag} respectively, while their {\it kinetic} and {\it magnetic} virial parameters follow the equations \eqref{eq:ak} and \eqref{eq:am}.

\subsection{Clump selection criteria} \label{sec:criteria}

From the clumps identified by the density thresholding procedure described in Sec.\ \ref{sec:sample} we retain only those that satisfy two additional criteria: First, the mass in sinks within the clump must be less than 10\% of the clump's gaseous mass. This criterion avoids considering clumps in which the gravitational potential is too strongly affected by the sinks' masses, and in which the feedback from the protostellar objects, not included in our simulation, should be taken into account in the energy budget. Second, we impose a minimum of 64 cells for each clump, which roughly translates to resolving the clump with 4 cells per dimension. 

Note that these criteria tend to discard the densest clumps, which tend to be the smallest and least massive ones when a sequence of hierarchically-nested clumps is considered. This is because of their small size, and because a given sink mass constitutes a larger fraction of a less massive clump. However, in some of the plots where the existence of these clumps is important, we show them with an explicit warning that they do not satisfy the selection criteria.

\section{Results}
\label{sec:results}

In this section, we analyze the data of the four regions, first together, as a single complex, and then separately, as individual clouds, to highlight the energy budget at different star formation activity levels. In the latter case, we can interpret each region as being in a different evolutionary stage \citep[e.g.,] [] {VS+09, VS+18, Zamora+14}.

\subsection{Scalings in the whole cloud complex} \label{sec:cl_complex}

\begin{figure}
	\includegraphics[width=\columnwidth]{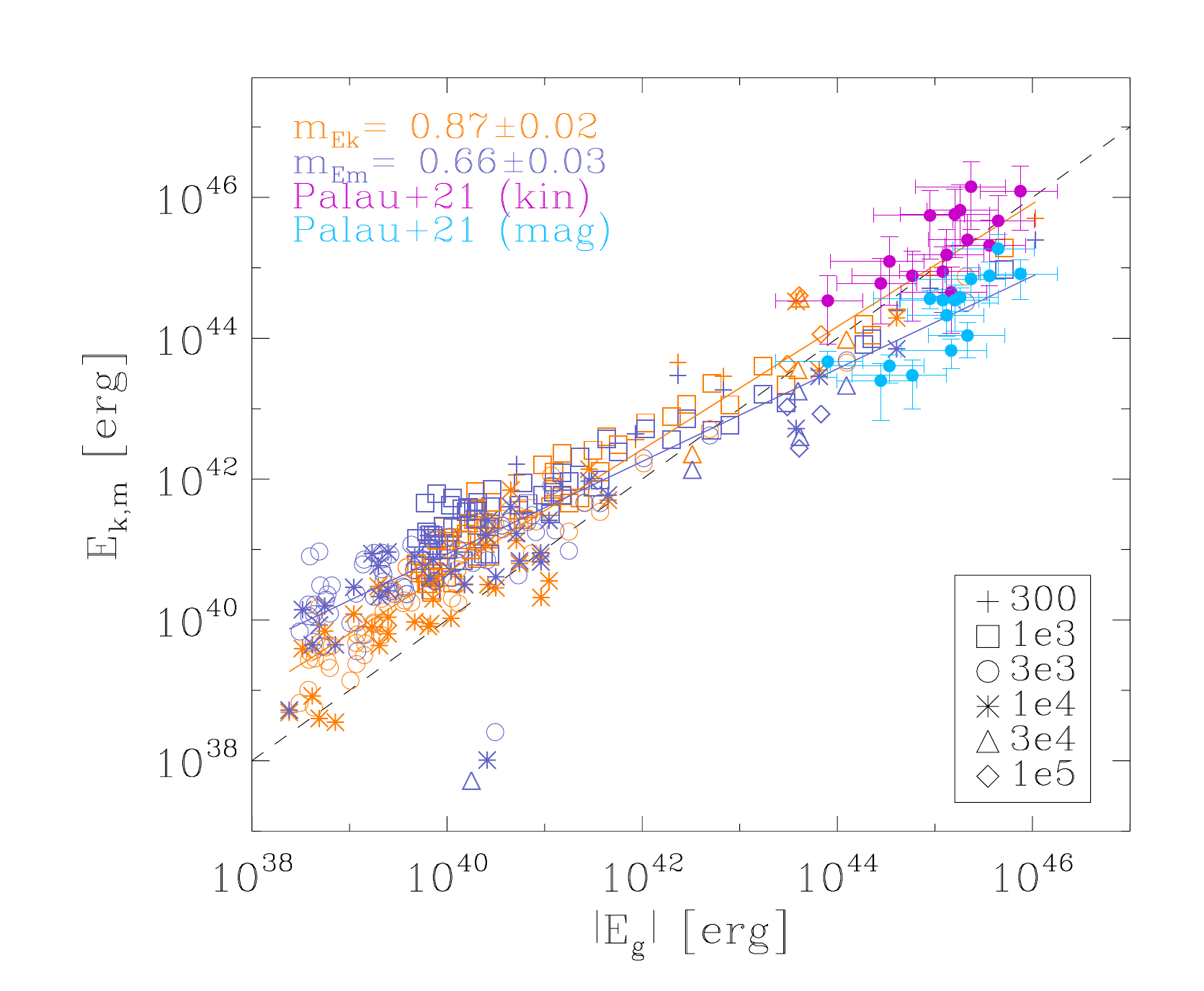}
    \caption{Kinetic energy $\Ek$ (orange) and magnetic energy $\Em$ (blue) versus gravitational energy $\Eg$ for clumps and cores in the simulated filament. Symbols represent the density threshold used for clump selection in units of $\pcc$. The dashed line separates the magnetically subcritical and supercritical regimes. The filled circles with error bars denote the observational core sample from \citet{Palau+21}, processed as described in Sec.\ \ref{sec:obs}.
    }
   \label{fig:energies}
\end{figure}

The clump sample in the present study is part of a larger collapsing GMC---a hub-filament system. To investigate how the magnetic field affects the energy budget of the structures in this region, we first compare their kinetic and magnetic energies to their gravitational energy in Fig. \ref{fig:energies}. 
In this figure, the different symbols denote the density thresholds used for the clump-finding algorithm in units of $\pcc$, while the colors denote the various quantities, with orange corresponding to the kinetic energy and blue to the magnetic energy. Linear fits in log-log are plotted for each case with their corresponding values and color.
In addition, the filled circles correspond to the dense cores from the observational sample in \citet{Palau+21}, in the magnetic (blue) and kinetic (red) cases. We discuss the comparison with the latter work and make reference to other observational samples in Section \ref{sec:obs}.

We can observe in this plot that the magnetic energy is comparable to the kinetic energy, and that both are  larger than $\Eg$, for clumps with low $\Eg$, while  $\Em < \Ek$ in clumps with large $\Eg$. In fact, it is seen that the two energies scale differently with $\Eg$, as indicated by the fits to the two sets of data points in this plot. Specifically, we find
\begin{equation}
\Ek \sim \Eg^{0.87\pm 0.02},
\label{eq:Ek_vs_Eg}
\end{equation}
while
\begin{equation}
\Em \propto \Eg^{0.66\pm0.03},
\label{eq:Em_vs_Eg}
\end{equation}
indicating that $\Ek$ increases faster with $\Eg$ than $\Em$. However, both energies scale with exponents smaller than unity, so that, in clumps with stronger gravitational binding, $\Eg$ is progressively larger than either $\Ek$ or $\Em$. 

\begin{figure}
	\includegraphics[width=\columnwidth]{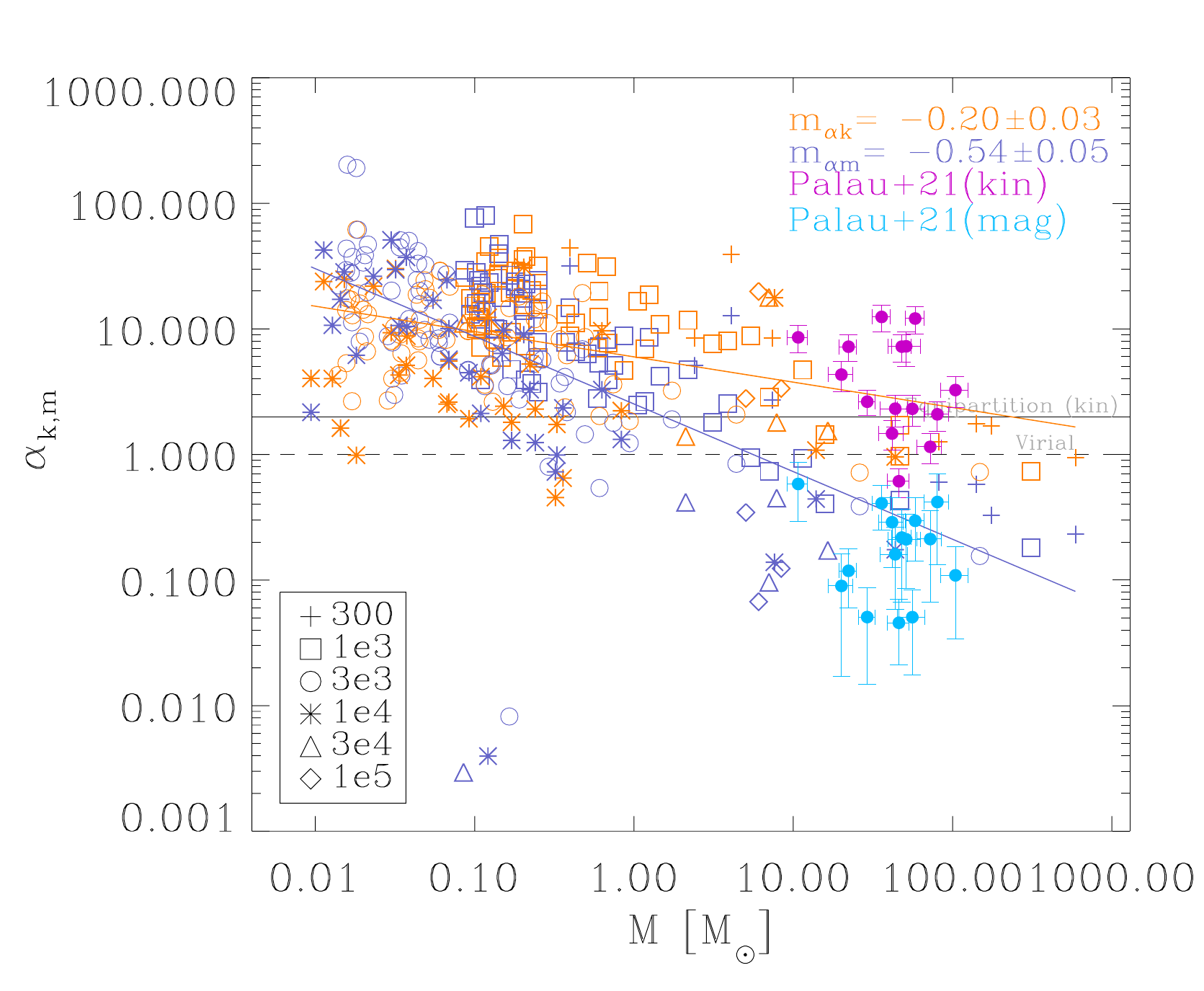}
    \caption{Virial parameter for the nonthermal kinetic energy $\akin = 2\Ek/|\Eg|$ (orange) and for the magnetic energy $\amag = \Em/|\Eg|$ (blue). Symbols correspond to the density threshold for clump selection, in units of $\pcc$; the solid line corresponds to equipartition (or free-fall) condition, and the dashed line to the virial condition. As in Fig.\ \ref{fig:energies}, the filled circles with error bars denote the observational core sample from \citet{Palau+21}, processed as described in Sec.\ \ref{sec:obs}.}
   \label{fig:alphas}
\end{figure}

The above result is also manifest in Fig.\ \ref{fig:alphas}, which shows the kinetic and magnetic ``virial parameters'' against clump mass. The kinetic and magnetic virial parameters are denoted by $\akin$ (orange symbols) and $\amag$ (blue symbols), and defined by eqs.\ \eqref{eq:ak} and \eqref{eq:am}, respectively.
The solid line corresponds to the equipartition or free-fall condition, $\Ek=|\Eg|$, in the kinetic case, and the dashed line to the kinetic virial equilibrium condition, $2\Ek = |\Eg|$, and to both equipartition and virial equilibrium in the magnetic case.
Indeed, it can be observed in Fig.\ \ref{fig:alphas} that both virial parameters decrease with increasing mass (corresponding to increasing gravitational energy), in agreement with the conclusion from Fig.\ \ref{fig:energies}. Moreover, $\amag$ is observed to decrease with $M$ faster than $\akin$, also consistent with the different slopes of $\Ek$ and $\Em$ with $\Eg$ seen in Fig.\ \ref{fig:energies}.

\begin{figure}
	\includegraphics[width=\columnwidth]{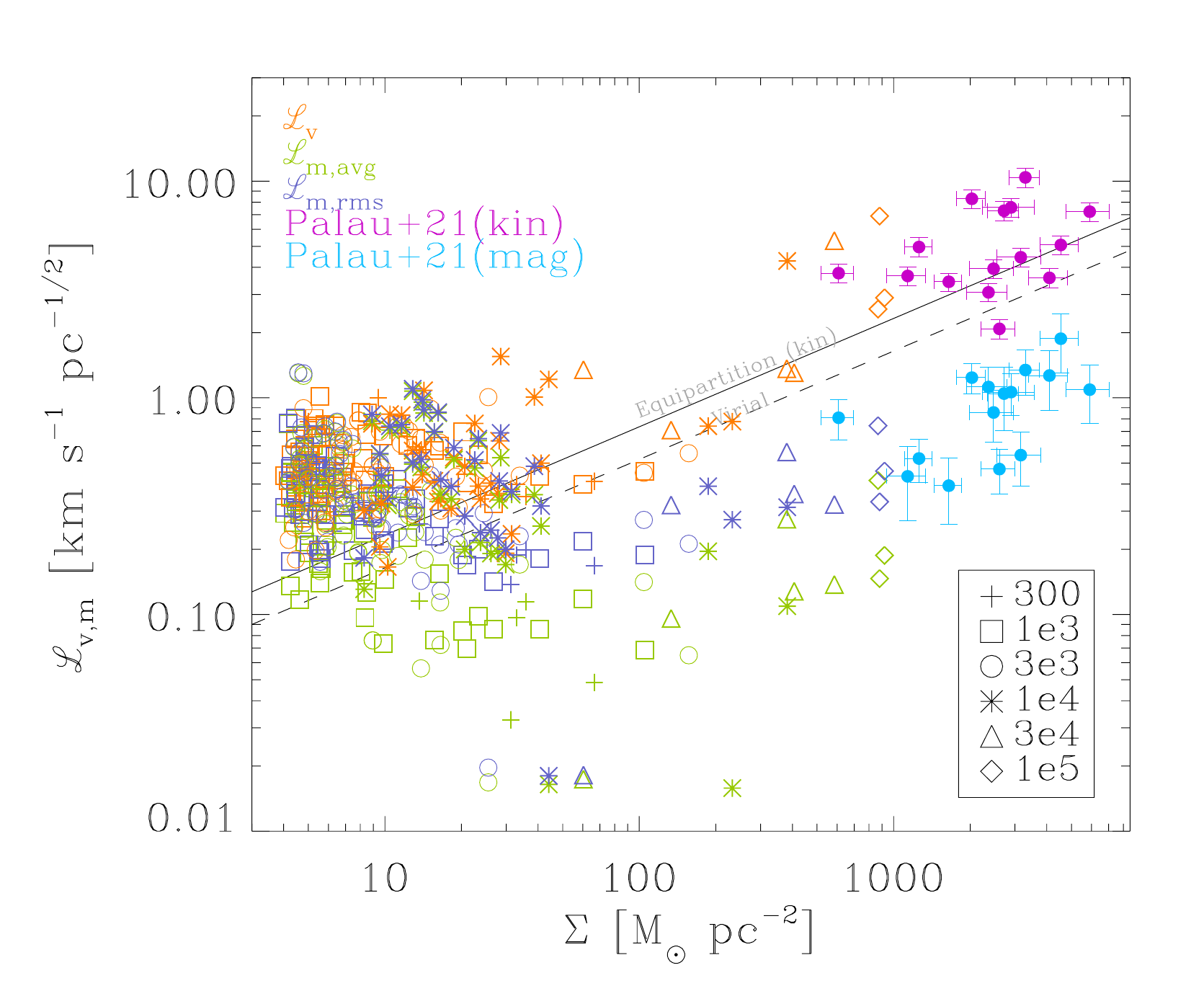}
    \caption{ Larson ratio for the clump sample denoting density thresholds with different symbols. $\Lcalv$ correspond to the classical Larson ratio (eq.\ [\ref{eq:lrat}], orange symbols). $\Lcalbmean$ and  $\Lcalrms$ correspond to the Larson ratio in the magnetic case (eq.\ [\ref{eq:lmag}]), considering either the mean magnetic field (green) or its rms value (blue). As in Fig.\ \ref{fig:energies}, the filled circles with error bars denote the observational core sample from \citet{Palau+21}, processed as described in Sec.\ \ref{sec:obs}.}
   \label{fig:lratios}
\end{figure}

The Larson ratios for the kinetic (orange symbols) and magnetic (blue and green symbols) energies are shown in Fig.\ \ref{fig:lratios}. Note that the blue and green symbols correspond to calculating the Larson ratio using either the {\it mean} or the {\it rms} value of the magnetic field strength in eq.\ \eqref{eq:lmag}, respectively.
Similarly to Fig. \ref{fig:alphas}, the solid and dashed lines respectively correspond to the free-fall (i.e., equipartition) and virial conditions. 

In agreement with the results of Paper~I, and of several observational works \citep[e.g.,] [] {Heyer+09, Leroy+15, Miville+17, Traficante+18a, Traficante+18b, Imara+19}, strongly supervirial values, with a large scatter, are observed at low-$\Sigma$ and low-$\nth$ in Fig.\ \ref{fig:lratios}. At intermediate $\Sigma$ values, the scatter is reduced and the sample becomes slightly sub-virial. Finally, at the largest column densities, there is a turnover that drives the trend over the equipartition region similarly to the results of \citet{BP+18} and \citet{Camacho+20}. Since this result is qualitatively identical to those from our previous nonmagnetic studies, we conclude that the magnetic field plays a mostly passive role in the evolution of the cores, without significantly altering the observed scaling of the kinetic virial parameter and Larson ratio. Moreover, since the clumps with the largest column densities in Fig.\ \ref{fig:lratios} again correspond to those with the largest gravitational energies, the trend seen in this figure is also consistent with that observed in Fig.\ \ref{fig:energies}.

Another important result is that the scalings of the magnetic virial parameter with mass, and of the Larson ratio with column density, are similar to those of their kinetic counterparts, in the sense that they both decrease for increasing mass/column density, albeit with slightly different slopes. This suggests that both the velocity and magnetic fields obtain their energy through similar mechanisms. We discuss this in Sec.\ \ref{sec:grav_driv_kin_mag}. 

Finally, it is worth noting that the values of the magnetic Larson ratio calculated using the mean field strength of the clump (green symbols) are systematically lower than those obtained using the rms value (blue symbols), indicating that a significant fraction of the magnetic energy is stored into the magnetic fluctuations within the clump, rather than in the mean field in each region.

\subsubsection{The $B$--$n$ relation} \label{sec:B-n_ensemble}

\begin{figure}
\includegraphics[width=\columnwidth]{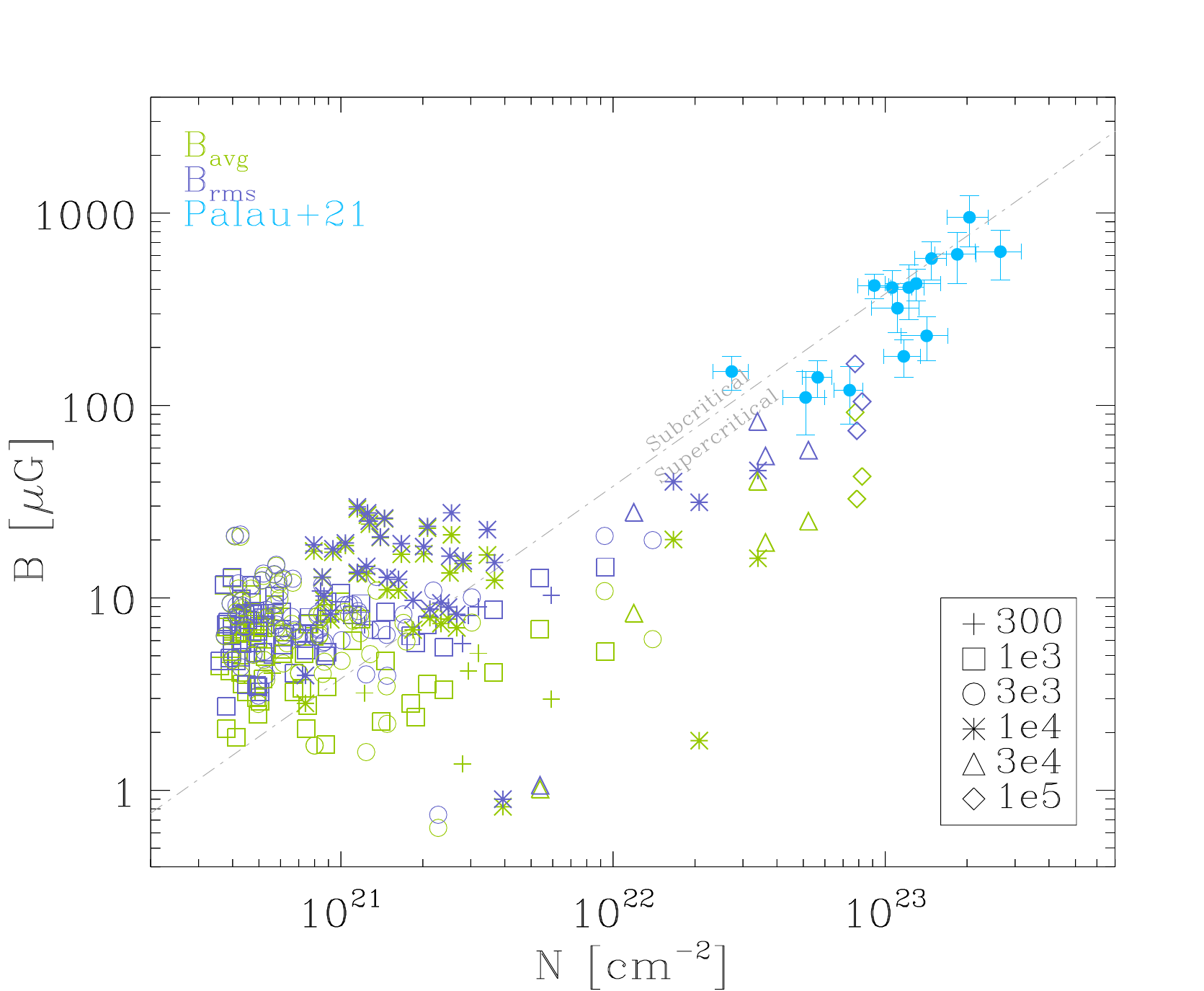}
    \caption{Magnetic field measurements, mean (green) and rms (blue), for the clump sample at different density thresholds ($\pcc$) against column density. As in Fig.\ \ref{fig:energies}, the filled circles with error bars denote the observational core sample from \citet{Palau+21}, processed as described in Sec.\ \ref{sec:obs}.}
   \label{fig:bfield}
\end{figure}

In Fig.\ \ref{fig:bfield} we show the mean and rms values of the magnetic field against column density. The dashed-dotted line separates the magnetically subcritical and supercritical regimes.  We observe that the upper envelope of the magnetic field strength remains roughly constant below $10^{22}$ cm$^{-2}$, while above this value, $B$ increases with increasing column density N(H$_2$) as well as with increasing $n$. Moreover, we note that the clumps become mostly magnetically supercritical for column densities $N(H_2) \gtrsim 3 \times 10^{21} \psc$, in agreement with the theoretical prediction of \citet{Hartmann+01} and the observational summary of \citet{Crutcher12} and \citet{Liu+22}. A direct comparison with the more recent compilation by \citet{Pattle+22} is not possible, since they do not plot the magnetic field against the column density. Finally, we compare our numerical results with our observational data sample (filled symbols with error bars in figure \ref{fig:bfield}) in Sec.\ \ref{sec:obs}.

\begin{figure*}
        \includegraphics[width=1.\columnwidth]{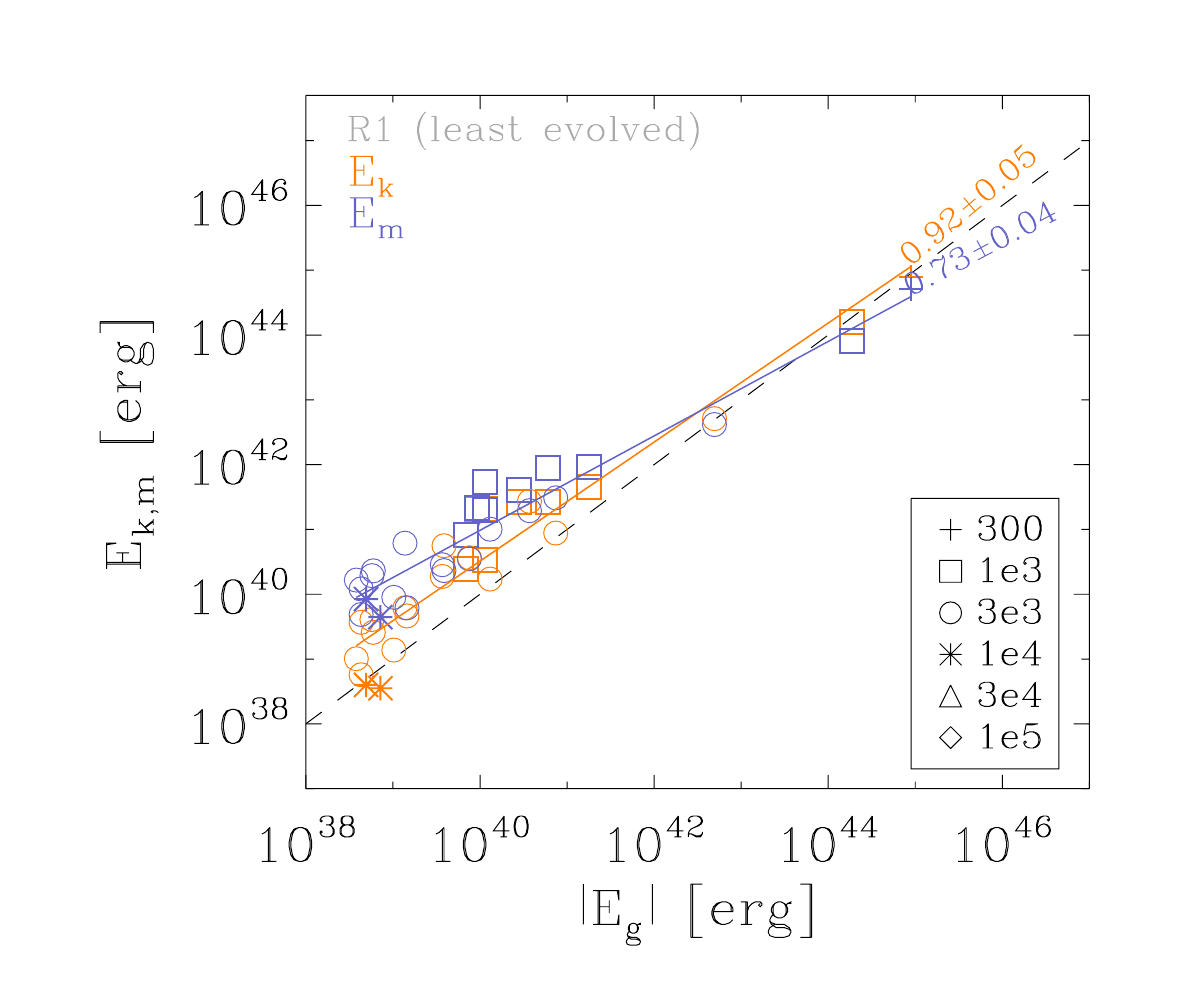}
        \includegraphics[width=1.\columnwidth]{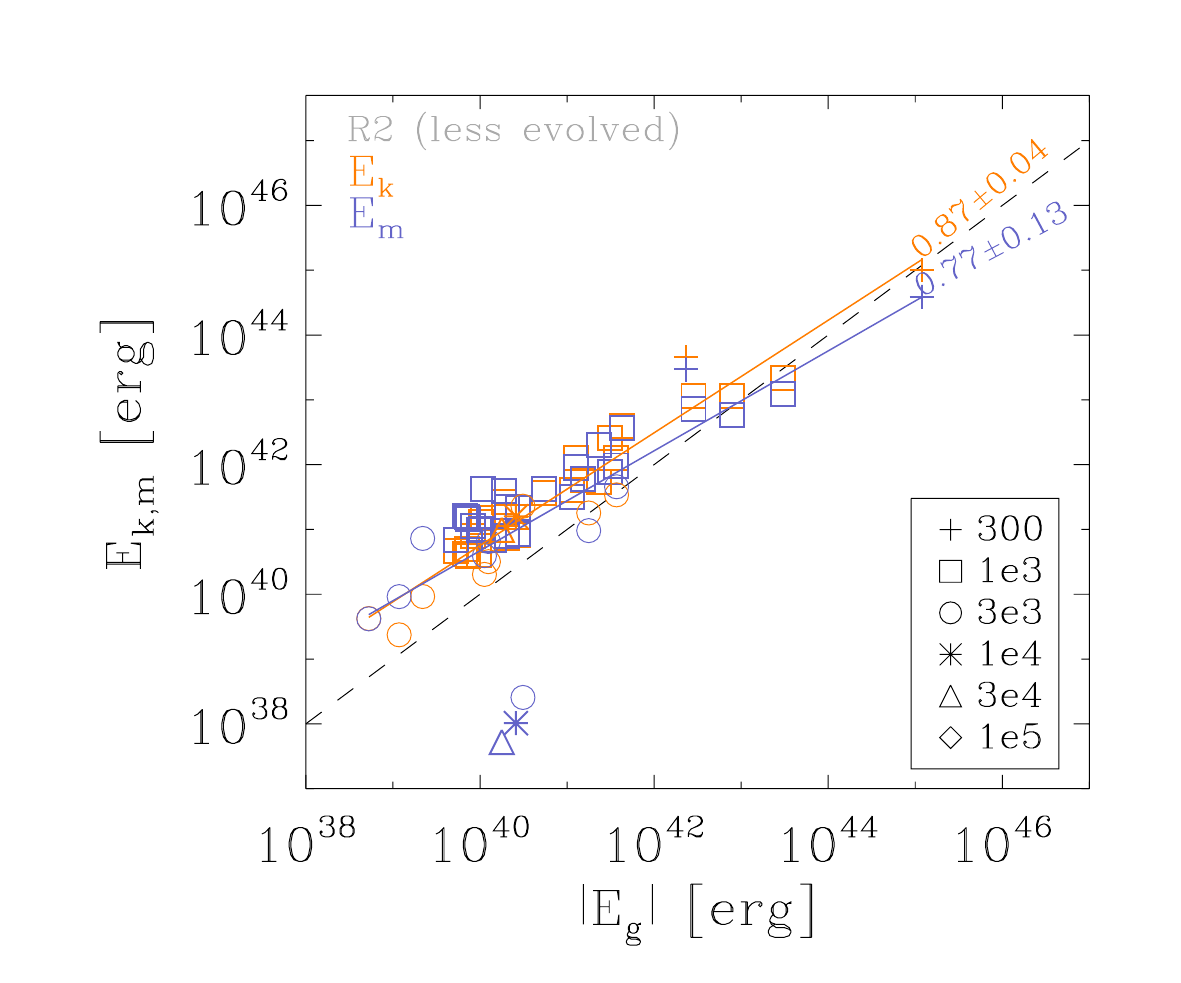} 
        
        \includegraphics[width=1.\columnwidth]{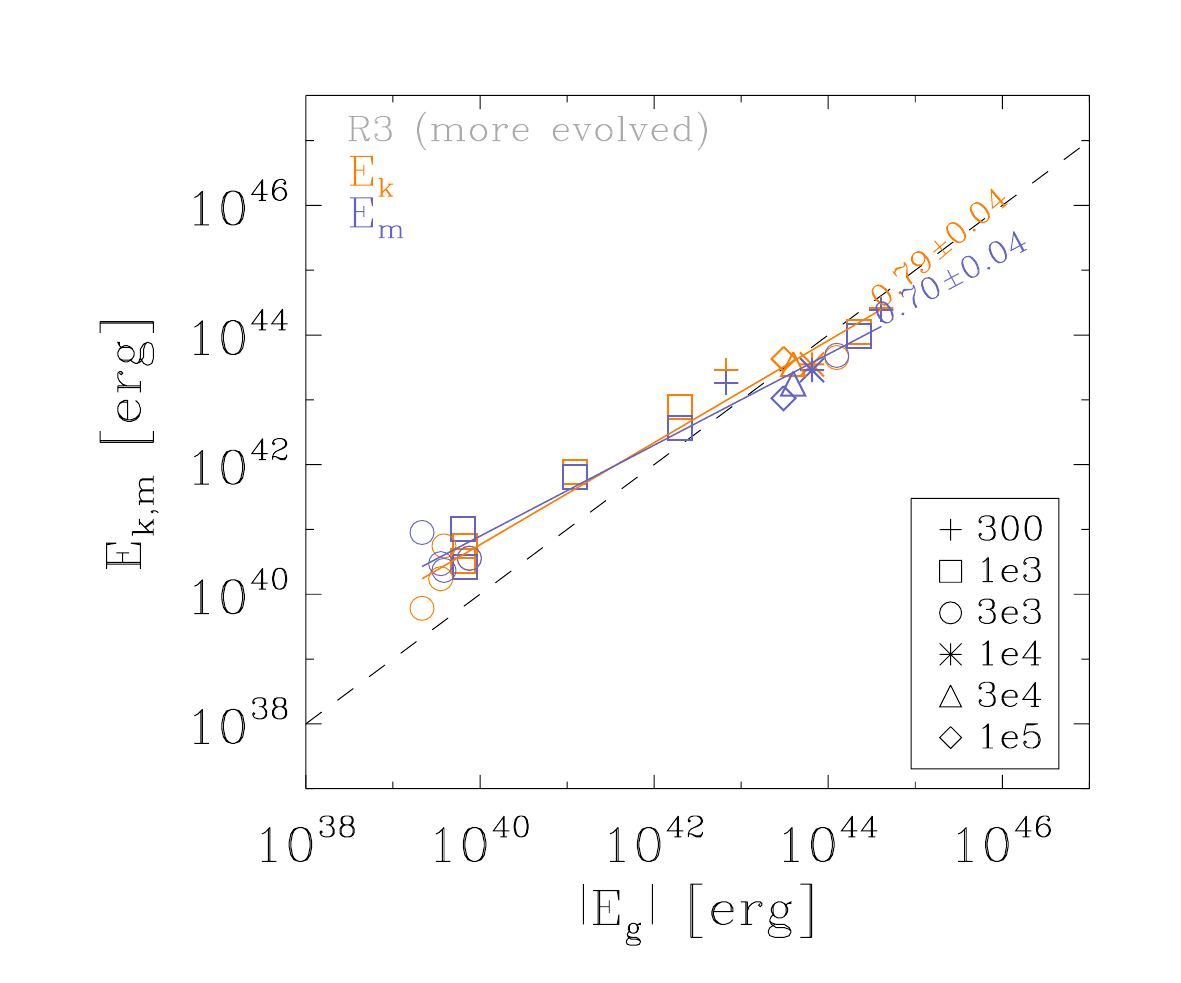}
        \includegraphics[width=1.\columnwidth]{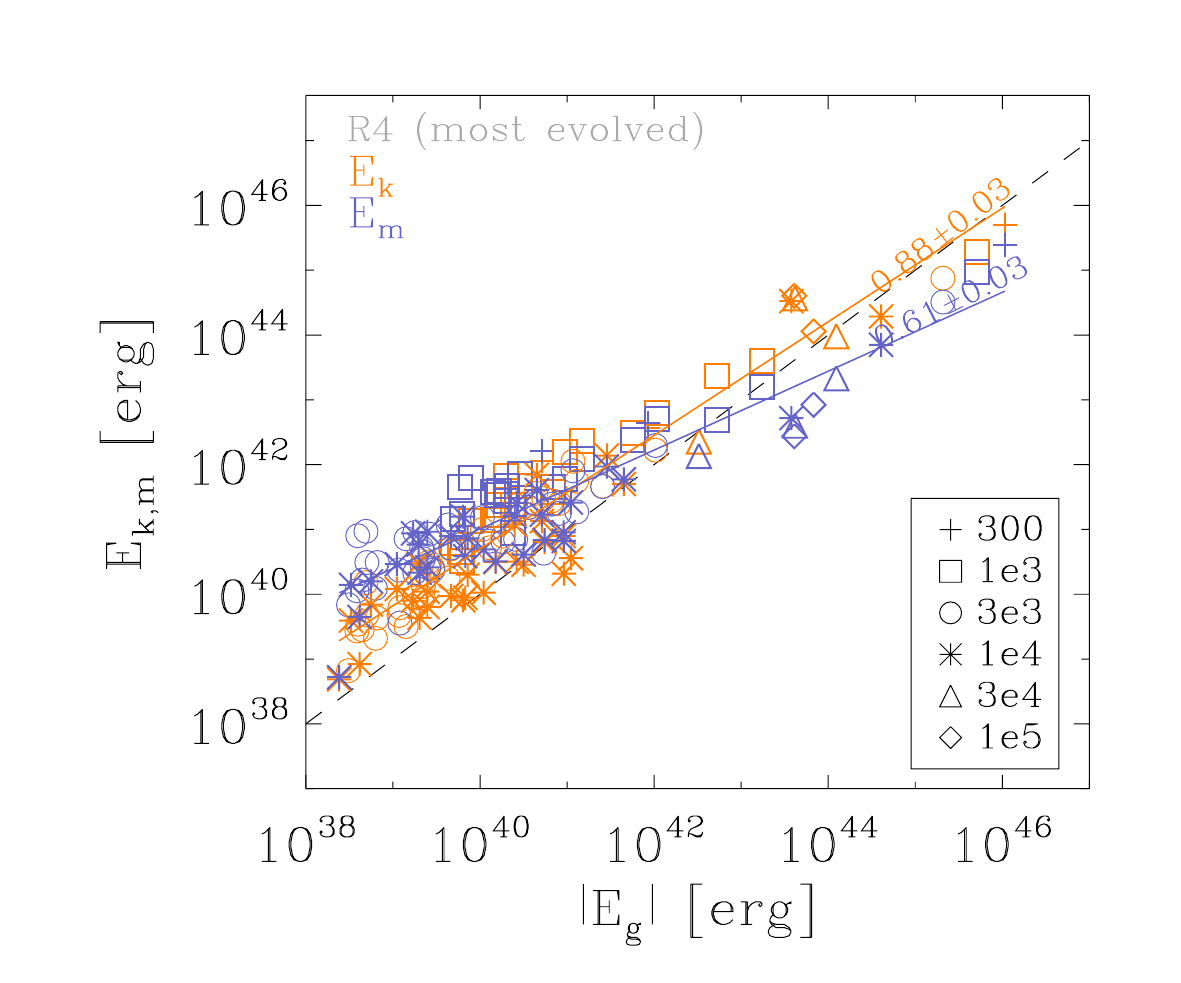}
    \caption{Comparison between kinetic and magnetic energies with gravitational energy for different regions of the simulated filament at $t=14.4$ Myr (see Fig. \ref{fig:sim}). Symbols correspond to the density threshold used for clump selection in $\pcc$. The regions are separated according to their age and star formation activity, which increase from R1 to R4.}\label{fig:energies_reg}
\end{figure*}

\begin{figure*}
        \includegraphics[width=1.\columnwidth]{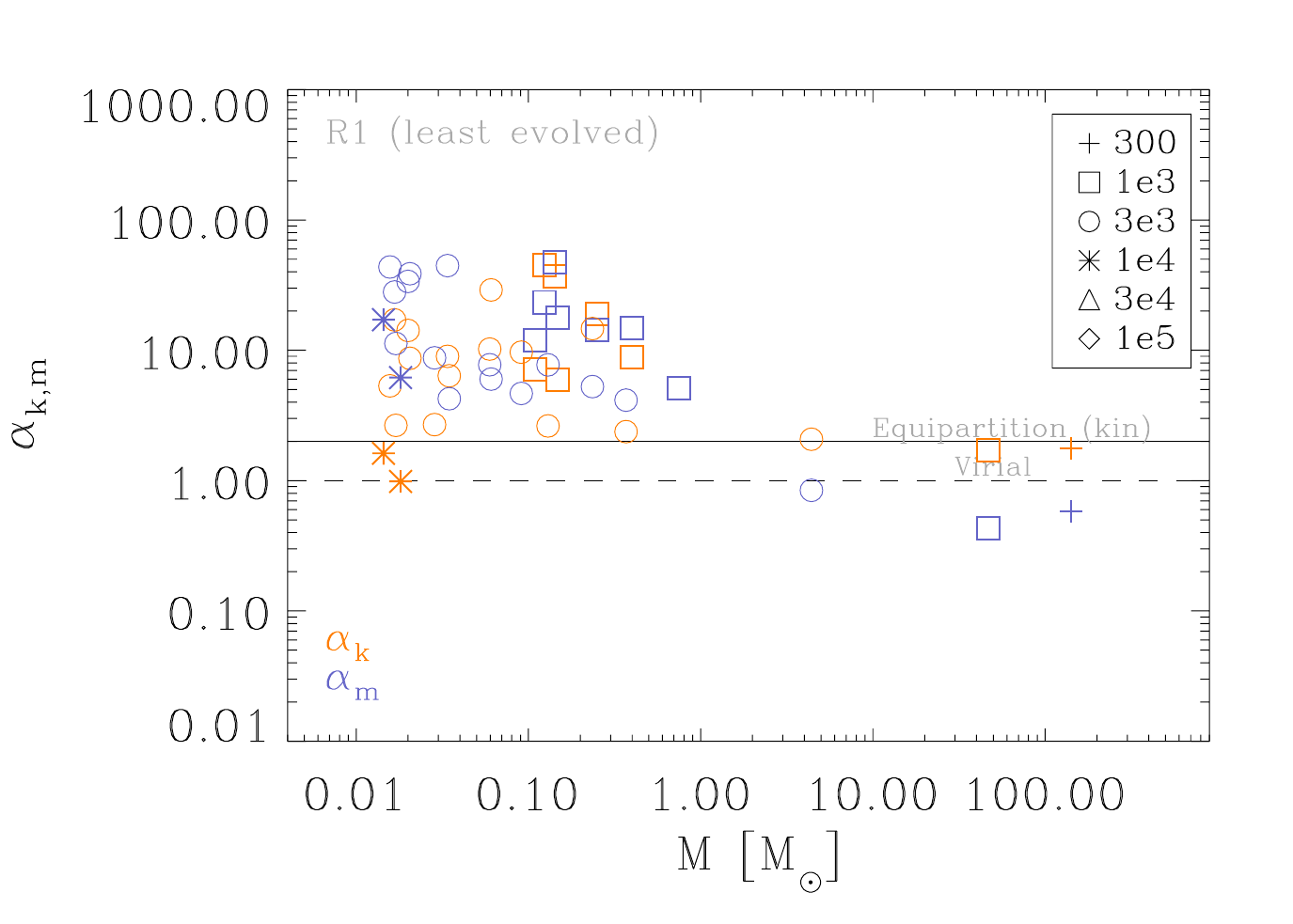}
        \includegraphics[width=1.\columnwidth]{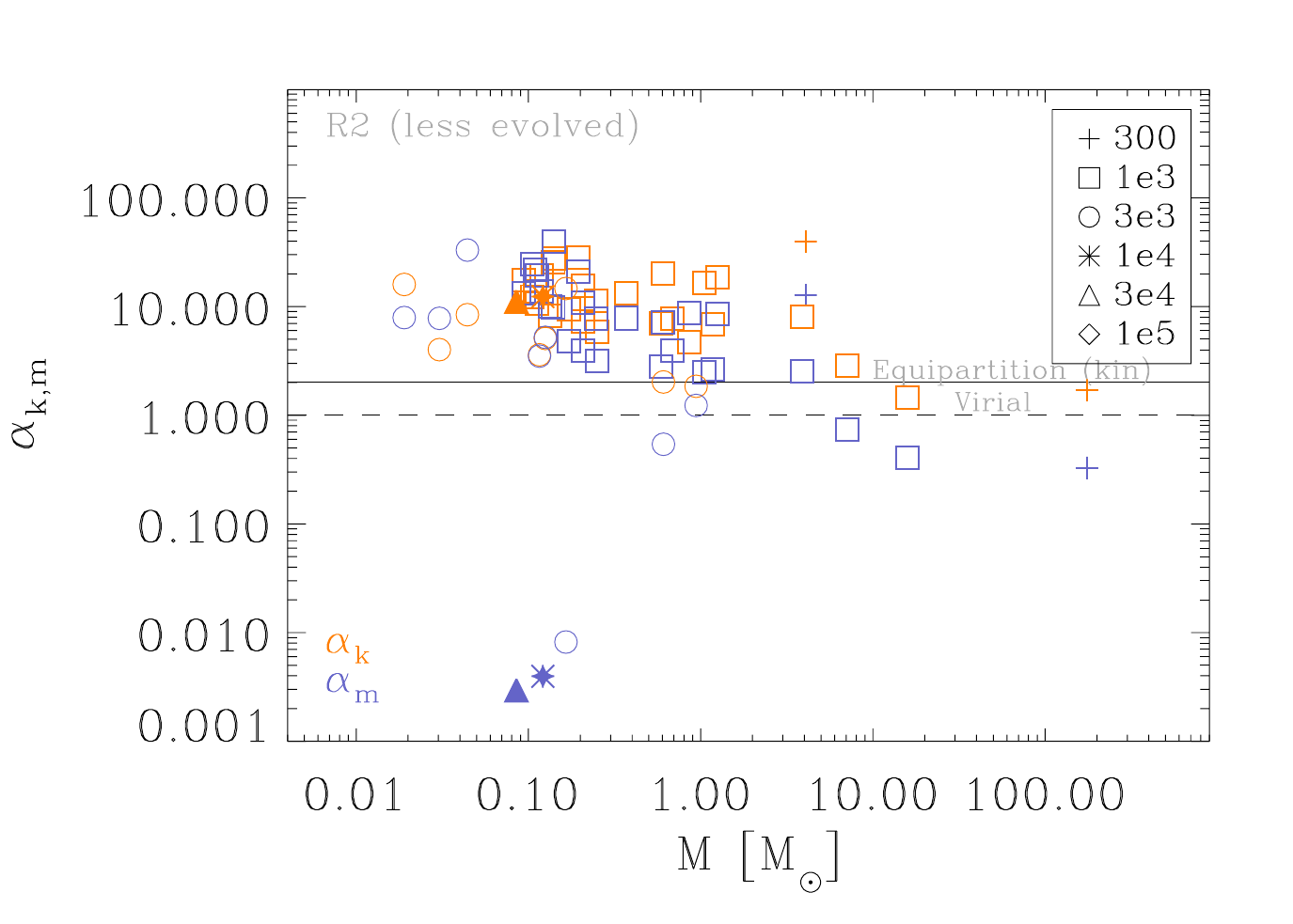} 
        
        \includegraphics[width=1.\columnwidth]{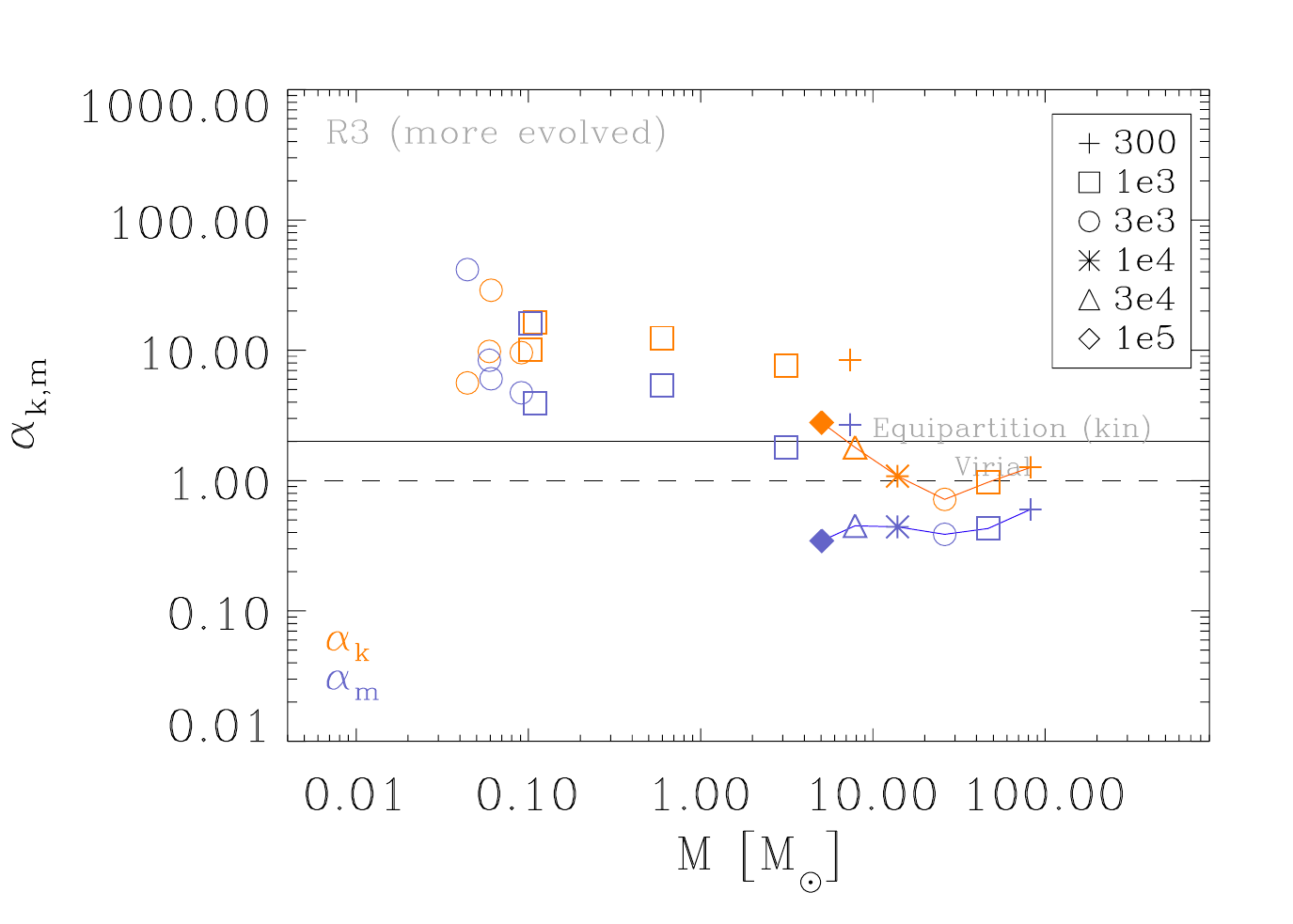}
        \includegraphics[width=1.\columnwidth]{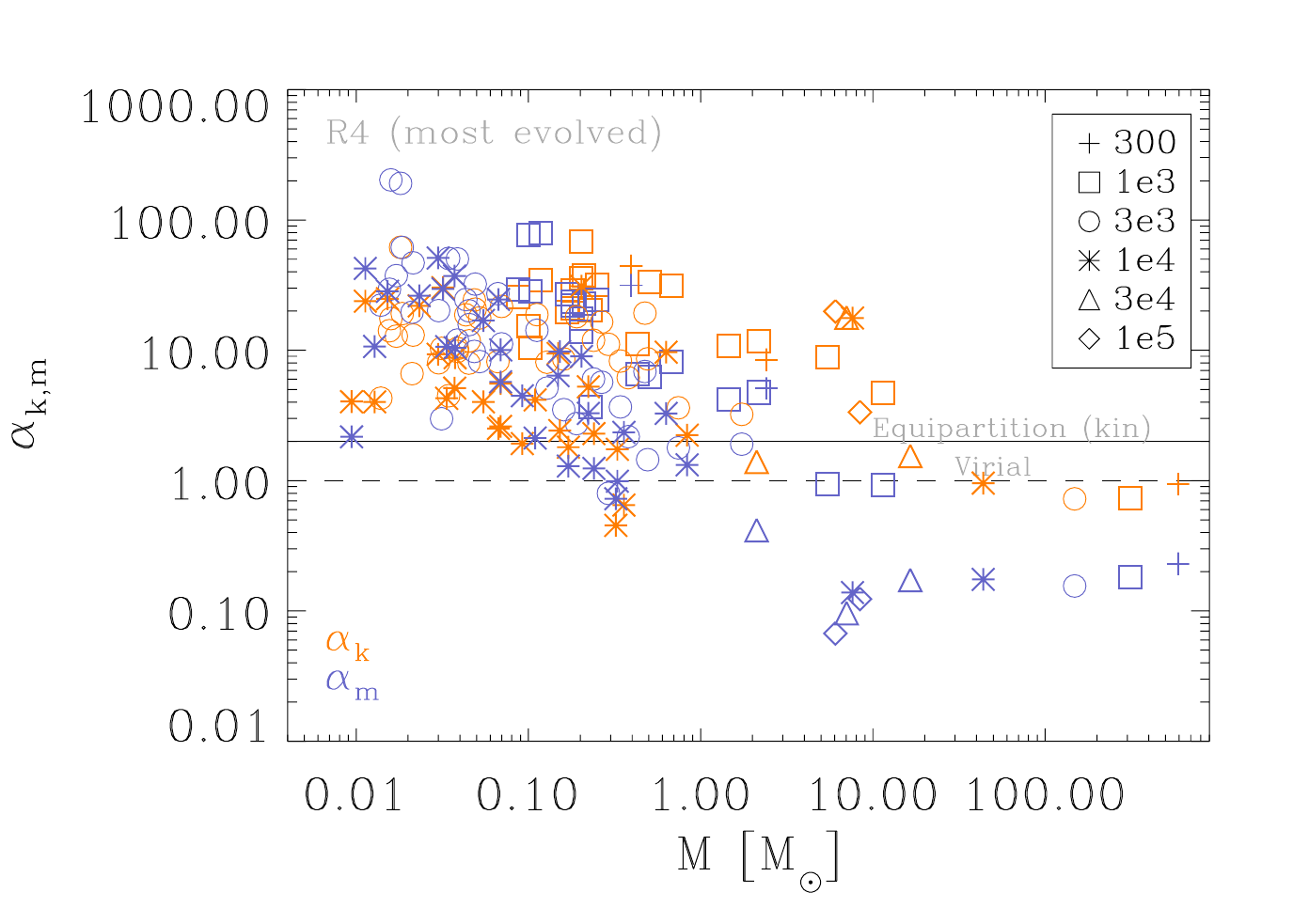}
    \caption{Kinetic and magnetic virial parameters in the four regions indicated by squares in the left panel of Fig.\ \ref{fig:sim} at t=14.4 Myr. In agreement with the model presented in \citet{VS+19}, the less evolved region, R1 with no sink particles, contains less massive objects with larger kinetic energies indicative of the external assembly phase, while the more evolved ones, R3 and R4, show a systematic decrease of $\alpha$ as the mass increases, except for a slight turnover at the highest masses. Note that the sequence of nearly virial points with $M \gtrsim 5 \Msun$ (from diamonds to``plus'' signs) in region R3 corresponds to a sequence of progressively larger, hierarchically-nested clumps within a single cloud. The filled symbols correspond to the densest cores containing a larger sink mass fraction than the 10\% selection criterion. In the bottom right panel, the line segments join a family of nested clumps whose $\akin$'s describe a concave curve, in agreement with the evolutionary study of Paper II.}
    \label{fig:alphas_reg}
\end{figure*}

\begin{figure*}
        \includegraphics[width=1.\columnwidth]{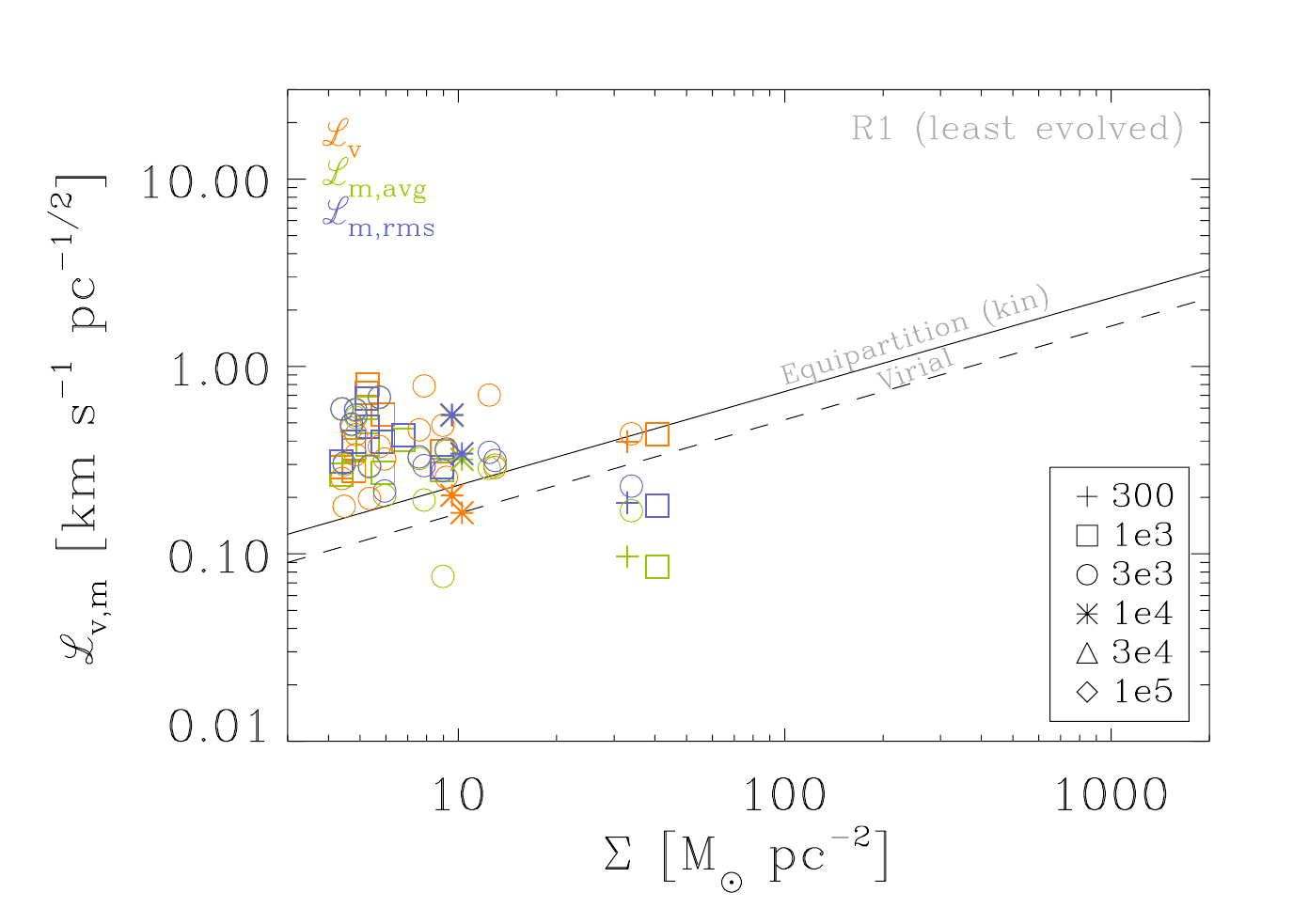}
        \includegraphics[width=1.\columnwidth]{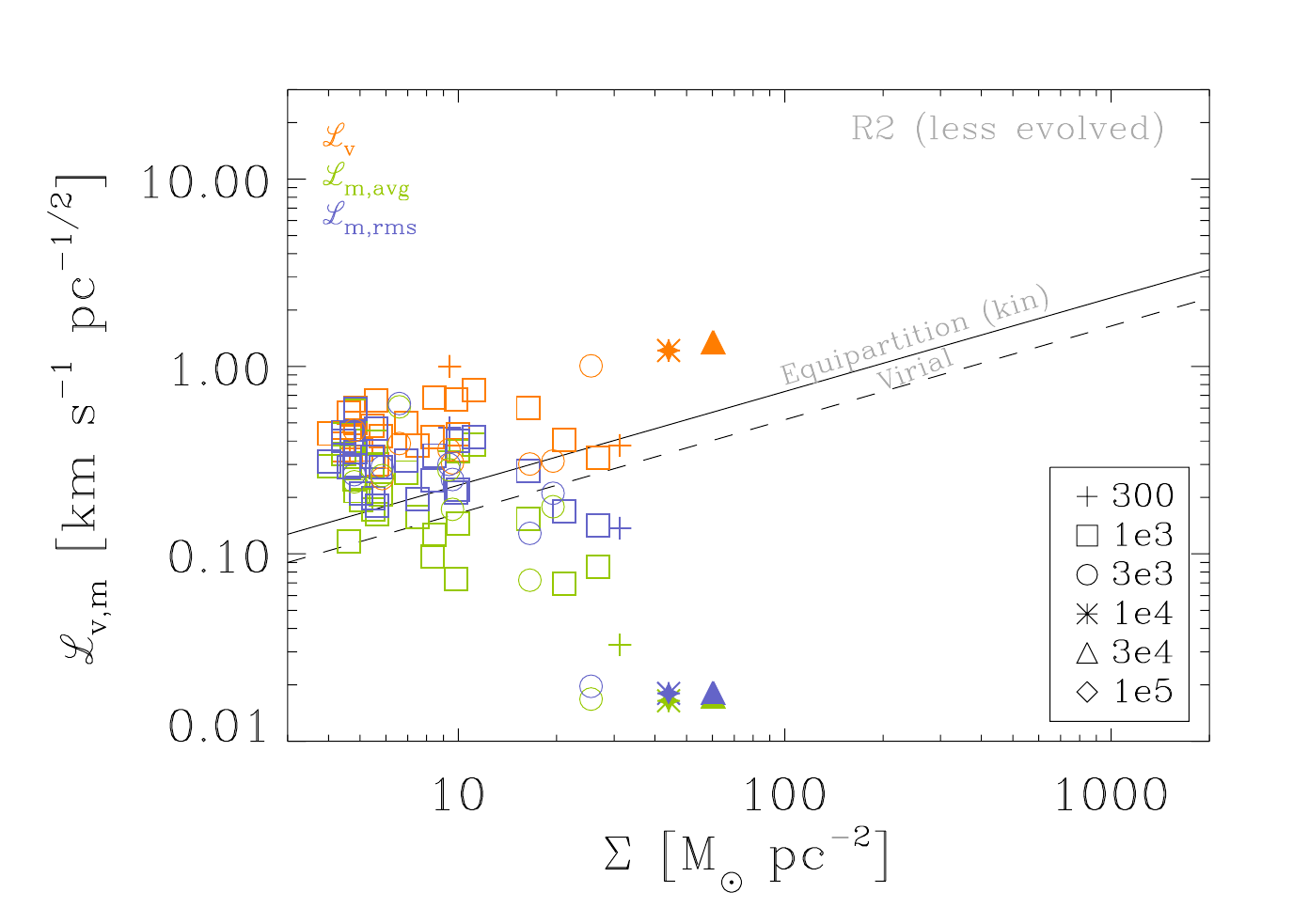} 
        
        \includegraphics[width=1.\columnwidth]{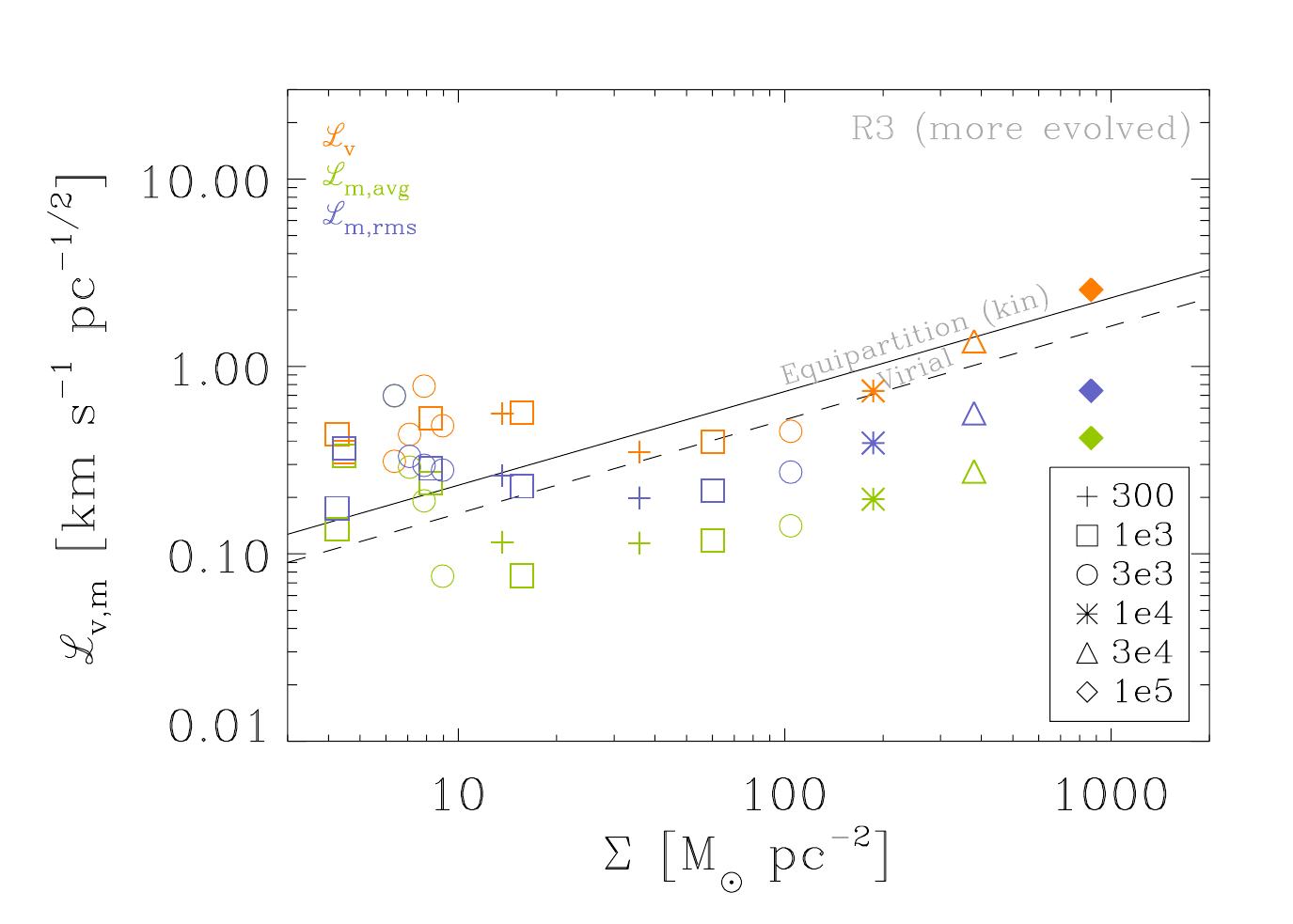}
        \includegraphics[width=1.\columnwidth]{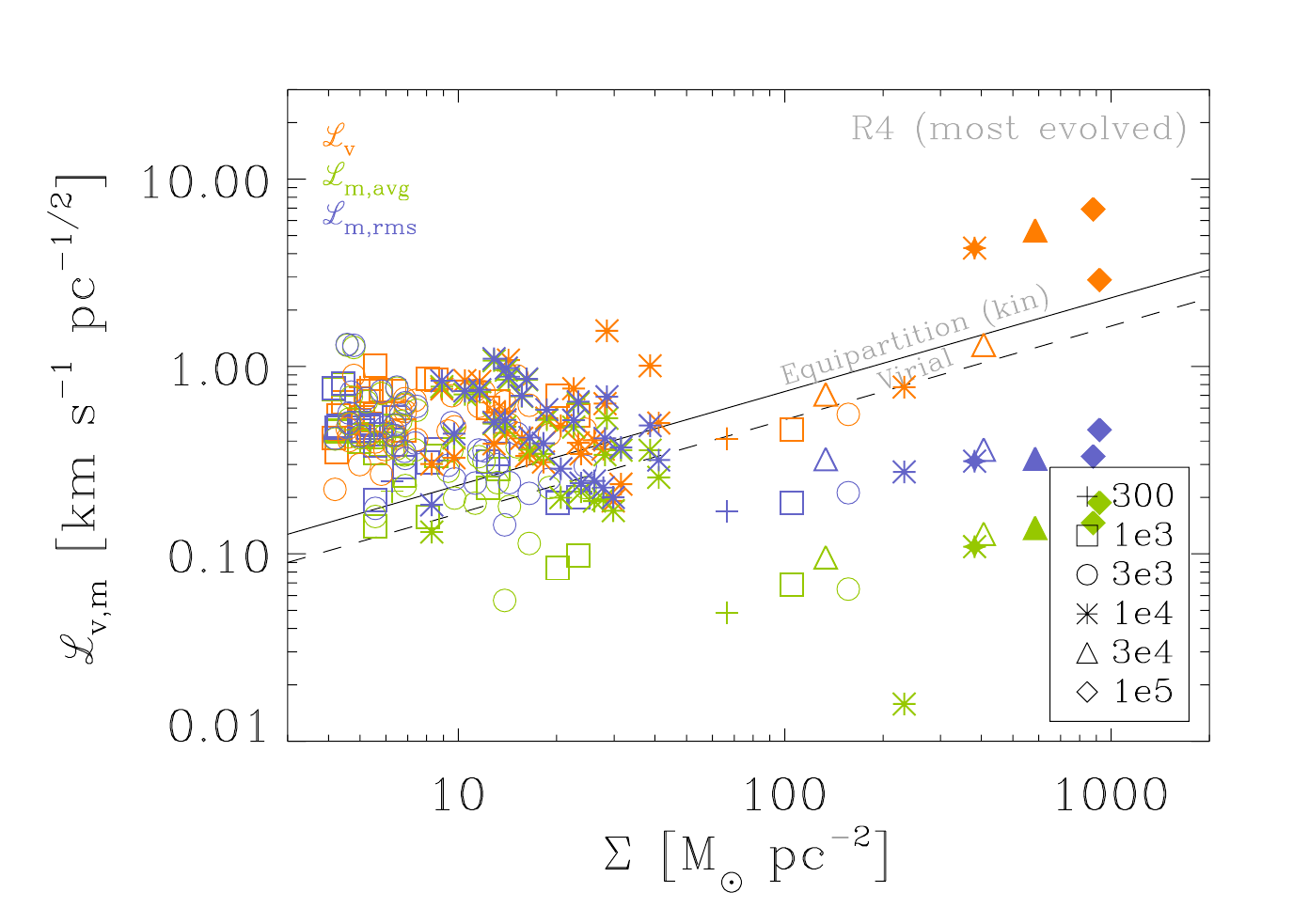}
    \caption{Kinetic and magnetic version of the Larson ratio for clumps in the regions indicated by squares in the left panel of Fig.\ \ref{fig:sim}, which constitute a sequence of progressively more evolved regions from R1 (top left panel) to R4 (bottom right panel). This sequence follows the evolution predicted in GHC model, which transits from a stage with an excess of kinetic energy and low column density, indicative of the external assembly phase, to a high-density stage in near equipartition at higher column density. The magnetic Larson ratio calculated with the rms magnetic field exhibits a similar qualitative behavior. Note that the sequence of points with $\Sigma \gtrsim 30\,\Msun$~pc$^{-2}$ (from ``plus'' signs to diamonds) in region R3 corresponds to a sequence of progressively more deeply embedded hierarchically-nested clumps within a single cloud. } 
    \label{fig:Lratios_reg}
\end{figure*}

\begin{figure*}
        \includegraphics[width=1.\columnwidth]{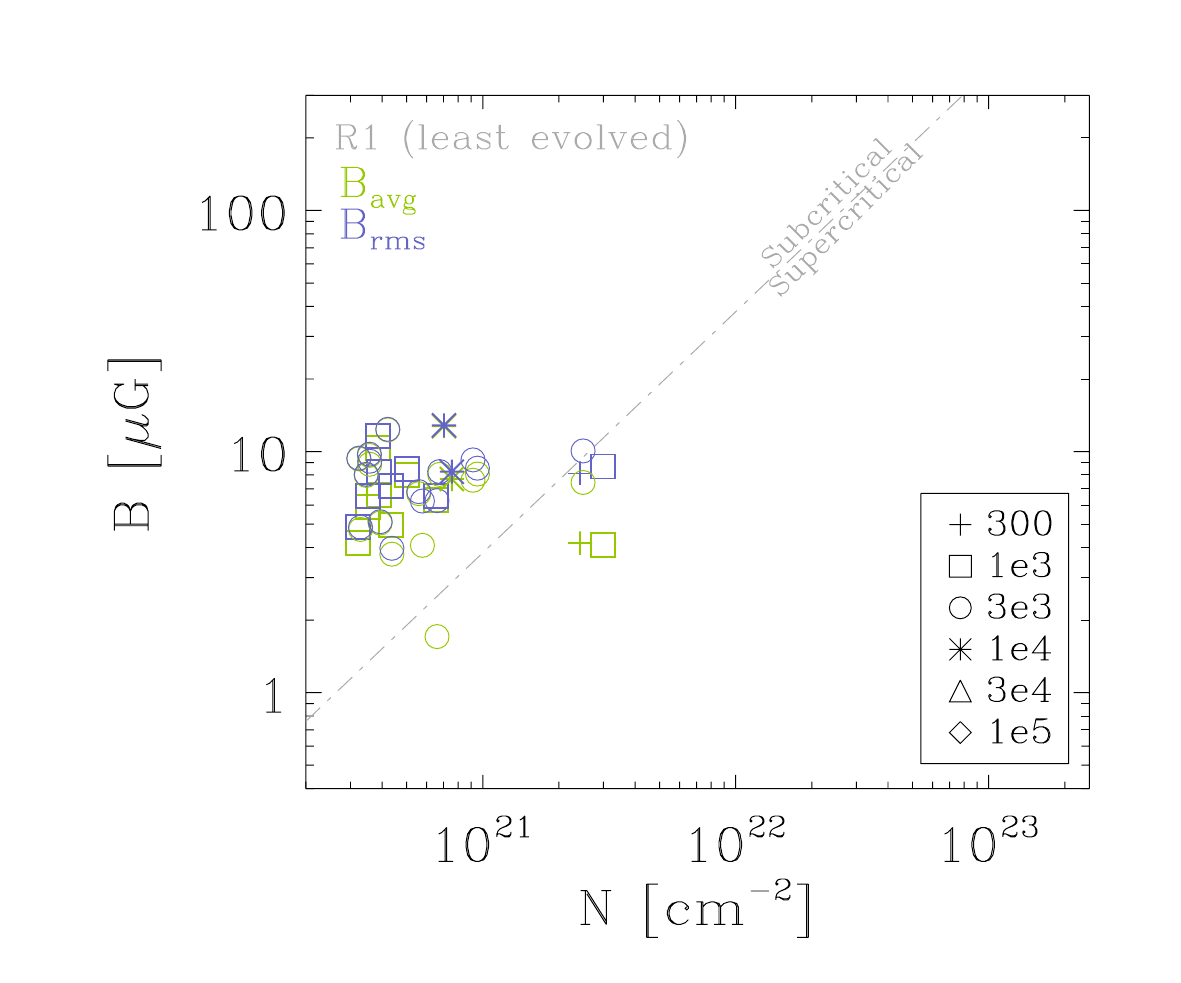}
        \includegraphics[width=1.\columnwidth]{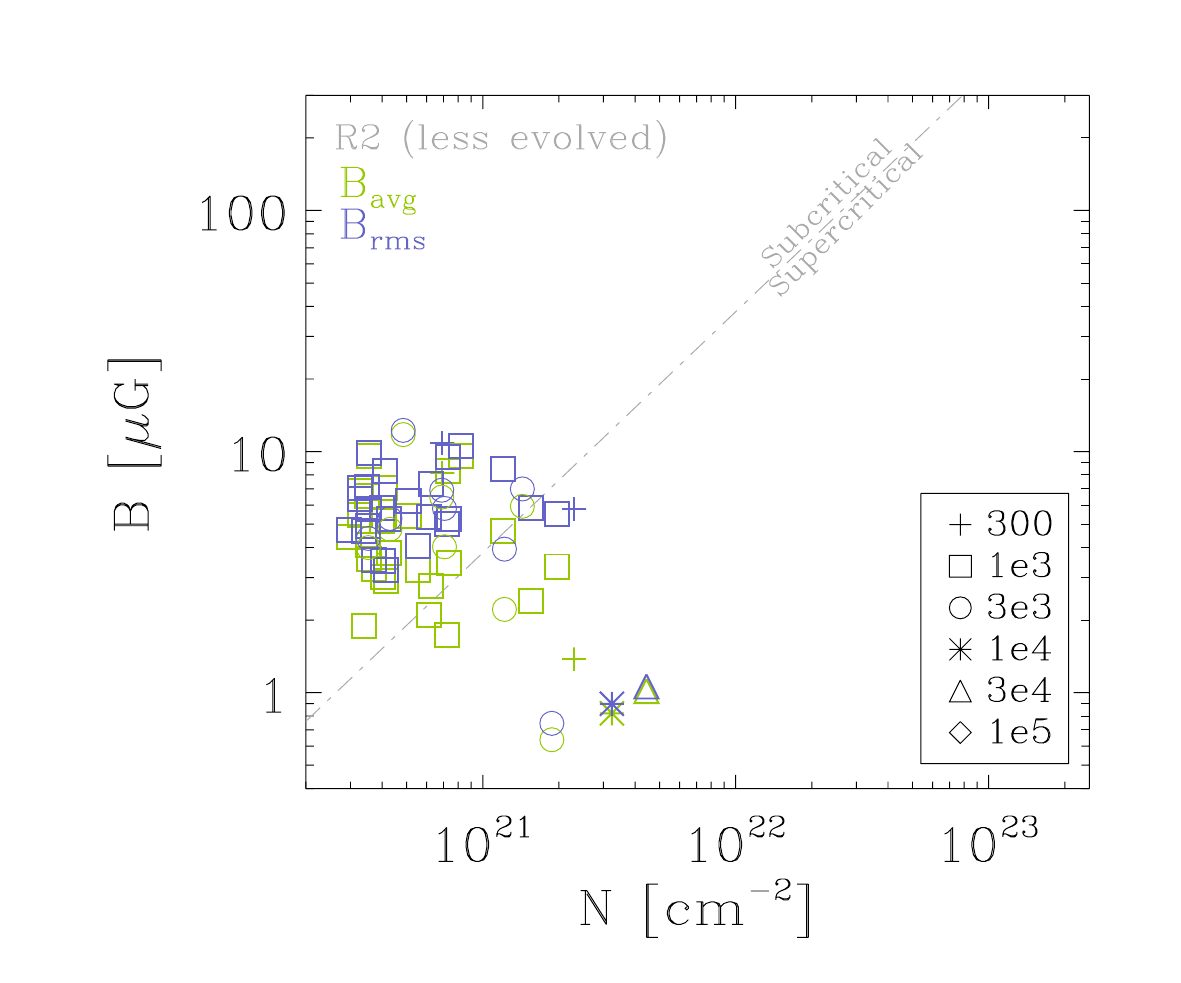} 
        
        \includegraphics[width=1.\columnwidth]{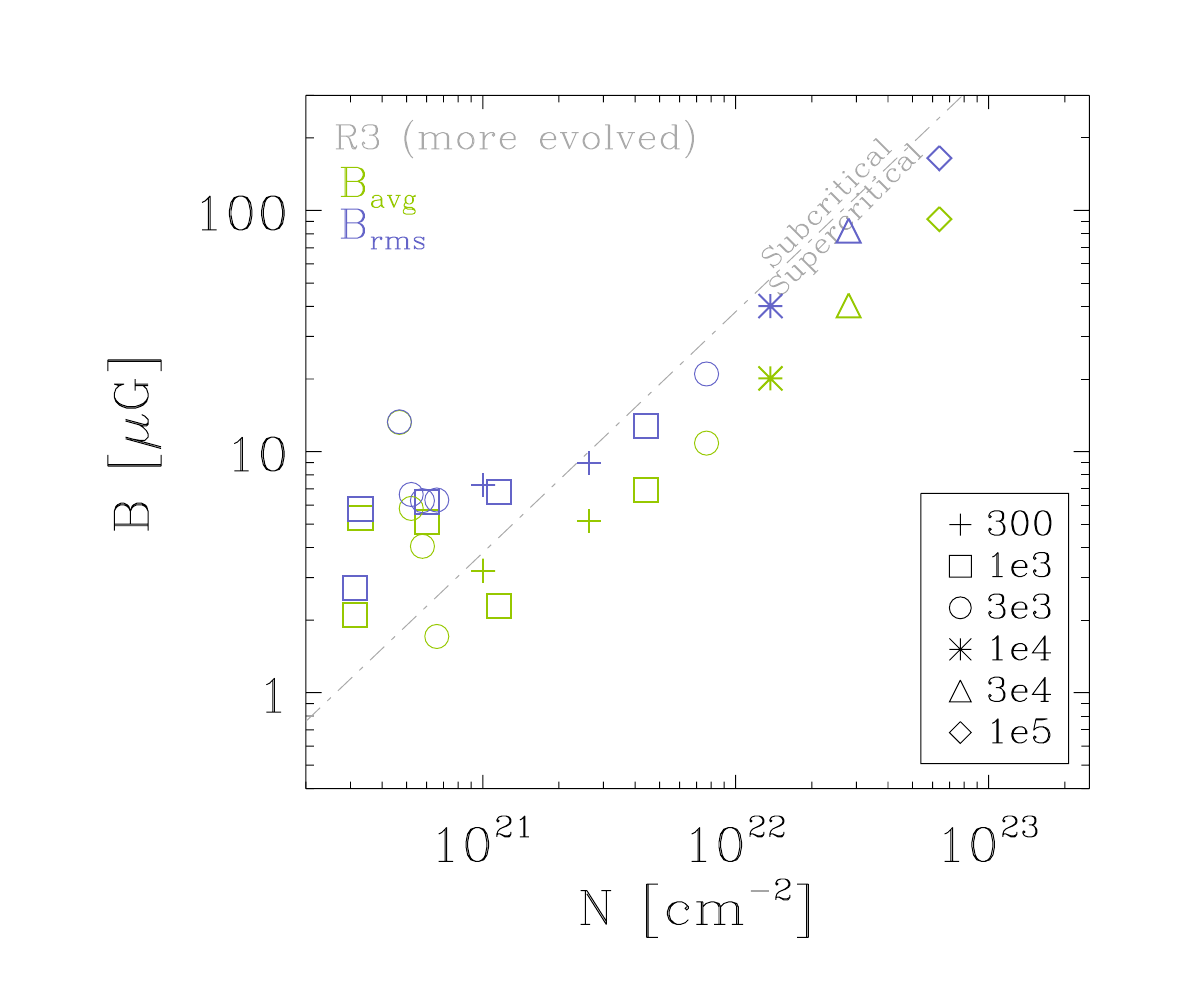}
        \includegraphics[width=1.\columnwidth]{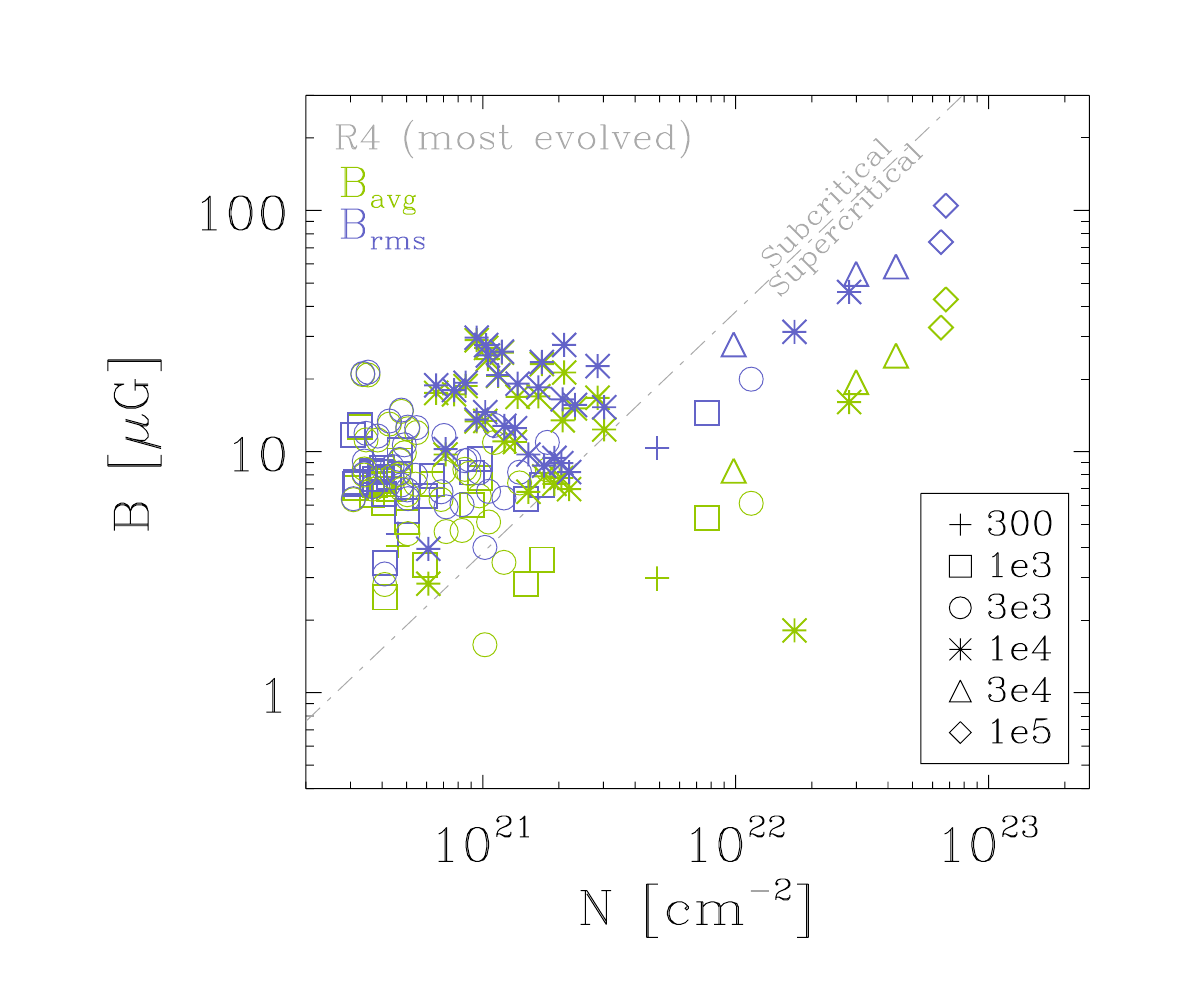}
    \caption{B versus column density $\mathrm{N}$, in the different regions. It can be observed the constant B field for low $\mathrm{N}$ values and the increment with a power law at larger $\mathrm{N}$ as suggested by \citet{Crutcher12}.}\label{fig:bsigmas_reg}
\end{figure*}

\subsection{Variation of the scalings across clumps  at different evolutionary stages} \label{sec:scaling_age}

In contrast with Figs.\ \ref{fig:energies}-\ref{fig:bfield}, which show the data for the full clump sample in the entire cloud, Figs.\ \ref{fig:energies_reg}-\ref{fig:bsigmas_reg} respectively correspond to regions R1-R4. As mentioned in Sec.\ \ref{sec:simulation}, these are characterized by increasing star formation levels from R1 to R4. Moreover, at $t=14.4$ Myr, regions R1-R3 in the filament have lower star formation than R4, which at that time, can be considered as the main hub. Therefore, the sequence of regions R1-R4 constitutes an evolutionary sequence.

Symbols and colors in Figs.\ \ref{fig:energies_reg}-\ref{fig:bsigmas_reg} have the same meaning as in figures \ref{fig:energies}-\ref{fig:bfield}. Figure \ref{fig:energies_reg} shows the kinetic and magnetic energies {\it versus} the gravitational energy for each of the four regions; Figure \ref{fig:alphas_reg} shows the kinetic and magnetic virial parameters {\it versus} mass; Figure \ref{fig:Lratios_reg} shows the kinetic and magnetic Larson ratios {\it versus} the column density $\Sigma$; and Figure\ \ref{fig:bsigmas_reg} shows the magnetic field strength {\it versus} $\Sigma$. 
We highlight the dense cores with high star formation efficiency with filled symbols in Figures \ref{fig:alphas_reg} and \ref{fig:Lratios_reg}. These dense cores do not comply with the selection criterion of having no more than 10\% of their mass in sinks; however, they were placed in the image to show that structures of these high densities do appear in the more evolved clouds.

In Figs.\ \ref{fig:energies_reg}-\ref{fig:bsigmas_reg}, we can observe that denser and more massive clumps progressively appear as the region considered is more evolved. Also, we note that the energy budget parameters $\alpha$ and $\Lcal$ tend to take subvirial values in those more massive or denser clumps.
These results are consistent with those of Paper II, where we showed that, as a clump evolves, it becomes more massive, develops higher density peaks, and increases its star formation rate (SFR). Simultaneously, we also showed that its budget parameters $\alpha$ and $\Lcal$ also evolve, describing concave tracks in the $\alpha$-$M$ and $\Lcal$-$\Sigma$ diagrams, starting out supervirial, then becoming noticeably subvirial, and finally approaching a near-virial value. 
In turn, these tracks are in qualitative agreement with the analytical model prediction presented in \citet{BP+18} and \citet{VS+19} for the evolution of the Larson ratio in the $\Lcal$-$\Sigma$ diagram. The agreement is only qualitative, because the precise trajectories in this diagram depend on the initial ratio of compressive kinetic energy to the clump's self-gravitating energy, and on the precise manner in which the compressive energy evolves during the clump's evolution. Nevertheless, the qualitative model prediction is consistent with an initial supervirial state, an intermediate subvirial one, and a final approach to a nearly virial state \citep[see Fig.\ 2 of] [] {VS+19}, thus generating a concave curve in the $\alpha$-$M$ and $\Lcal$-$\Sigma$ diagrams analogous to the one observed here. The clearest example is that of the locus of points in the bottom right panel of Fig.\ \ref{fig:alphas_reg}, which are joined by a segmented line, for identification.

\subsection{Scaling with depth within individual clumps at a fixed time} \label{sec:scaling_depth}

An important point to note in Figs. \ref{fig:energies_reg}-\ref{fig:bsigmas_reg} is that, within each of the individual regions R1-R4, many of the high-density clumps are actually the inner parts of the larger, lower-density ones defined in that region. Yet, the hierarchically nested objects occupy similar loci as those that are not nested. For example, consider the three orange square symbols in the mass range $4 < M/\Msun <20$ in the upper right panel (region R2) in Fig.\ \ref{fig:alphas_reg}. Being clumps defined at the same density (indicated by the square symbol), they correspond to different clumps. These can be compared to the sequence of nested objects indicated by the star-triangle-diamond sequence of orange symbols in the same mass range in the panel corresponding to region R3 (bottom left panel of Fig.\ \ref{fig:alphas_reg}). These objects are all part of the same cloud, and correspond to a sequence of progressively more deeply embedded and denser regions within it.

We observe that the nested objects in R3 span roughly the same locus in the $\alpha$-$M$ diagram as the distinct clouds of similar masses in R2. This means that the virial parameter varies with depth within a cloud in a similar manner as it varies among different objects of similar masses that are presumably at different evolutionary stages. We discuss this further in Sec.\ \ref{sec:equiv_time_depth}.

\section{Discussion and implications}
\label{sec:disc}

\subsection{The gravitational energy as the unifying parameter} \label{sec:dom_Eg}

The trends in both the \am\ (Fig. \ref{fig:alphas}) and the \LS\ (Keto-Heyer, Fig. \ref{fig:lratios}) diagrams, in both the kinetic and magnetic cases, are seen to be summarized in the energy diagrams, of $\Ek$ and $\Em$ {\it vs.} $\Eg$ (Fig. \ref{fig:energies}). In fact, it is seen that there is much less scatter in these energy diagrams than in either the $\alpha$-$M$ or the Keto-Heyer diagrams. 

This result can be understood as follows: in the \am\ diagram, the most massive objects, which are also the largest, clearly have strong self-gravity, $\Eg$. However, low-mass objects can have either large or small self-gravitational energies depending on whether they are small and dense, or large and diffuse, respectively. Analogously, in the \LS\ diagrams, the high-column density objects clearly have large $\Eg$, but the low-$\Sigma$ objects can have either large or small $\Eg$ depending on whether they have large or small sizes/masses, respectively. On the other hand, sorting the objects directly by their self-gravitational energy eliminates the ambiguity, indicating that the true relevant parameter in determining $\alpha$ and $\Lcal$ (in both their kinetic and magnetic versions) is precisely $\Eg$. That is, both the kinetic and magnetic energy content of the clumps are mostly of gravitational origin.

Moreover, the energy diagrams for the individual regions (Fig.\ \ref{fig:energies_reg}) show that the younger, less evolved regions (R1 and R2) have fewer objects of large $\Eg$, and reach lower values of it, while the number of high-$\Eg$ objects, and the maximum value of this energy, are larger in the more evolved objects. Since in the low-$\Eg$ objects, $\Ek$ and $\Em$ are often larger than $\Eg$, and the opposite occurs in the high-$\Eg$ objects, these results suggest that the low-$\Eg$ objects are still in the process of assembly by external agents, but evolve to become self-gravitating, and that this process occurs at all scales.

However, it is also crucial to note that the highest-$\Eg$ objects in each of the regions are always defined at the lowest densities (indicated by the plus signs). These are the parent objects in each region, and they are seen to be {\it always} dominated by their self-gravity. This implies that their substructures, even if they are not initially self-gravitating themselves, are being formed by the gravitational contraction of their parent structure. This is precisely the expectation for a multi-scale collapse regime in which the collapse begins at the largest scales, while the collapses of their internal substructures start later (although they culminate earlier) \citep{VS+19}.

\subsection{Temporal variation of $\alpha$ and $\Lcal$} \label{sec:evol_alpha_L}

We have seen in Sec.\ \ref{sec:scaling_age} that clumps in the younger regions (R1 and R2; top-left and top-right panels in Figs.\ \ref{fig:alphas_reg} and \ref{fig:Lratios_reg}) have supervirial values of the energy budget parameters $\alpha$ and $\Lcal$. This can be understood as a consequence of the fact that {\it all clumps must be born supervirial} since, after all, when the clumps are only starting to grow, their density is not much larger than that of their parent structure, and so their self-gravitating energy must be relatively small. The compression that forms them must have an external origin either inertial (such as a passing shock or a generic turbulent compression), or gravitational (such as the gravitational potential of a stellar spiral arm, or of the clump's parent gaseous structure). The kinetic energy in this external compression is thus larger than the clump's self-gravitating energy, causing its budget parameters to be large. Later, as the clumps become denser and more massive, they become locally self-gravitating, and their budget parameters, $\alpha$ and $\Lcal$, decrease.

In the particular case of our clump sample, the upper-left  panel in Fig.~\ref{fig:Lratios_reg} shows that the youngest region (R1) contains a population of mostly supervirial clumps of low column density. Nevertheless, these clumps must be undergoing compression because this region is already gravitationally bound when defined by thresholds of $\nth = 300$, 1000, and 3000 $\pcc$ (plus, square and circle symbols, respectively), and it is clearly denser 1 Myr later (compare the left and right panels of Fig.\ \ref{fig:sim}). This suggests that the compression acting on the individual clumps within this region is due to the self-gravity of the entire region. 
This is precisely the effect reported by \citet{Gomez+21}, that there is always a certain inner region (a fragment) within a larger collapsing region that is effectively non-self-gravitating, and is only contracting because it is being compressed by the infalling external material. In this case, the gravitational potential responsible for the contraction of the fragment is that of its parent structure. 

\begin{figure*}

\includegraphics[width=\columnwidth]{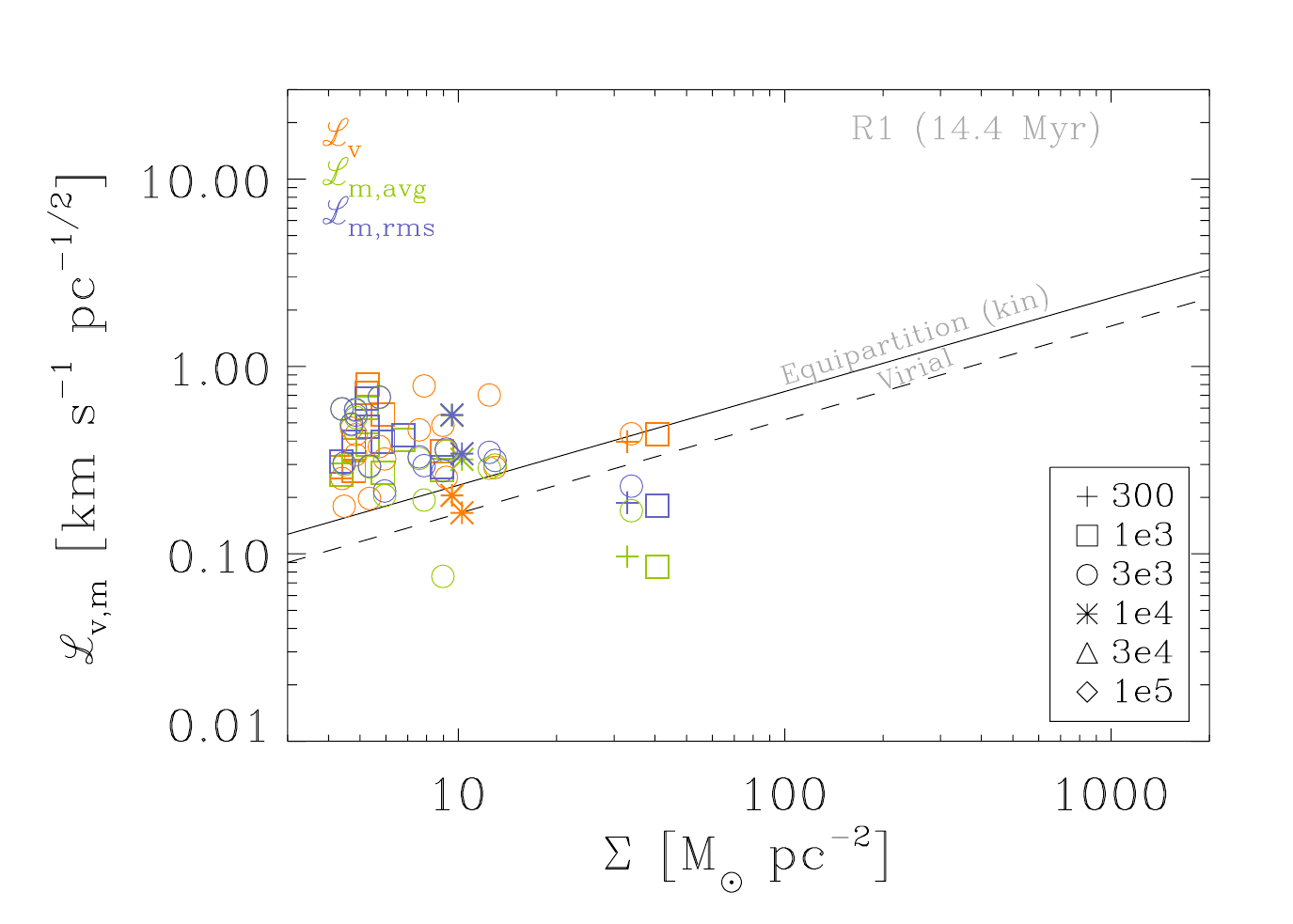}
\includegraphics[width=\columnwidth]{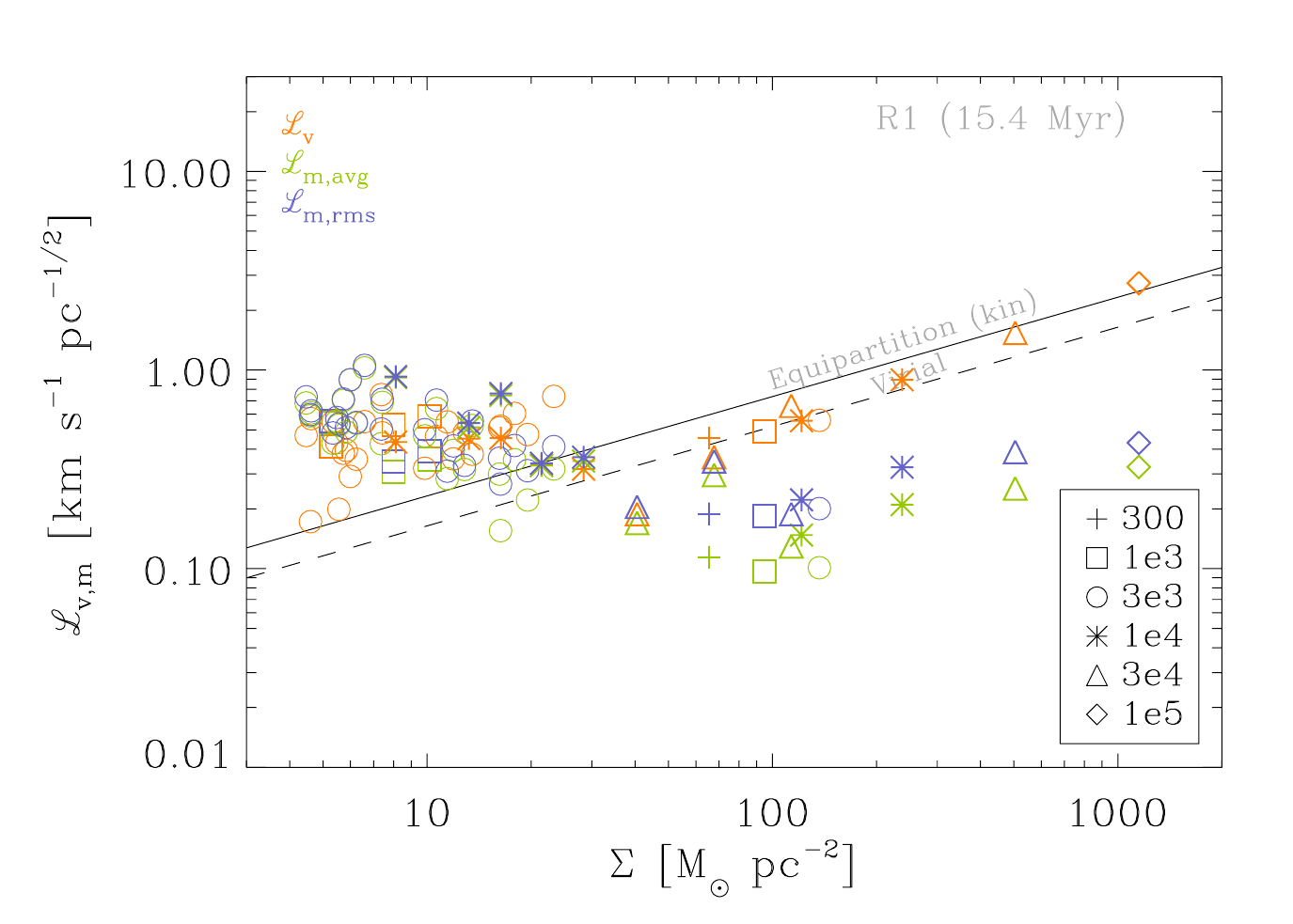}
\includegraphics[width=\columnwidth]{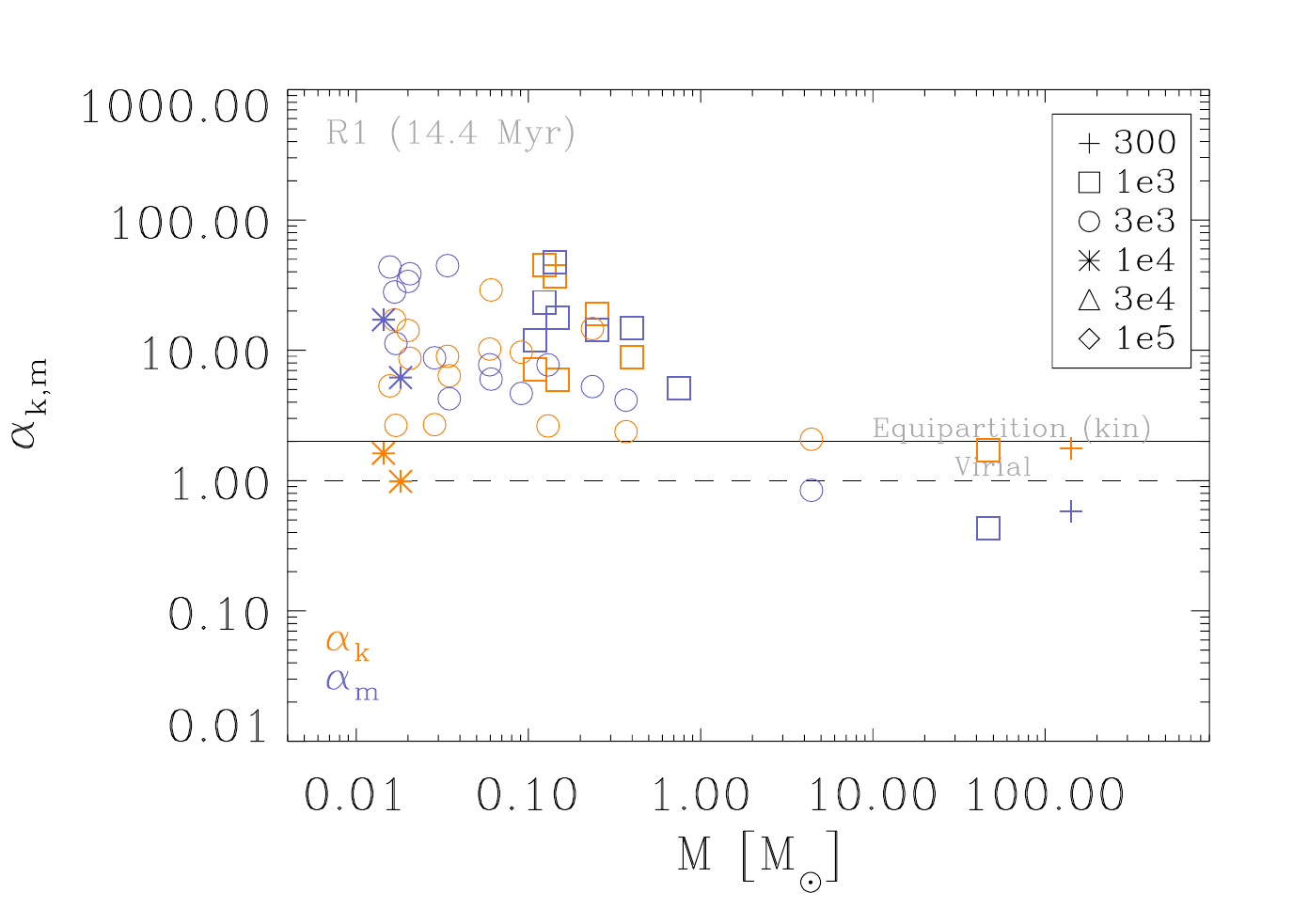}
\includegraphics[width=\columnwidth]{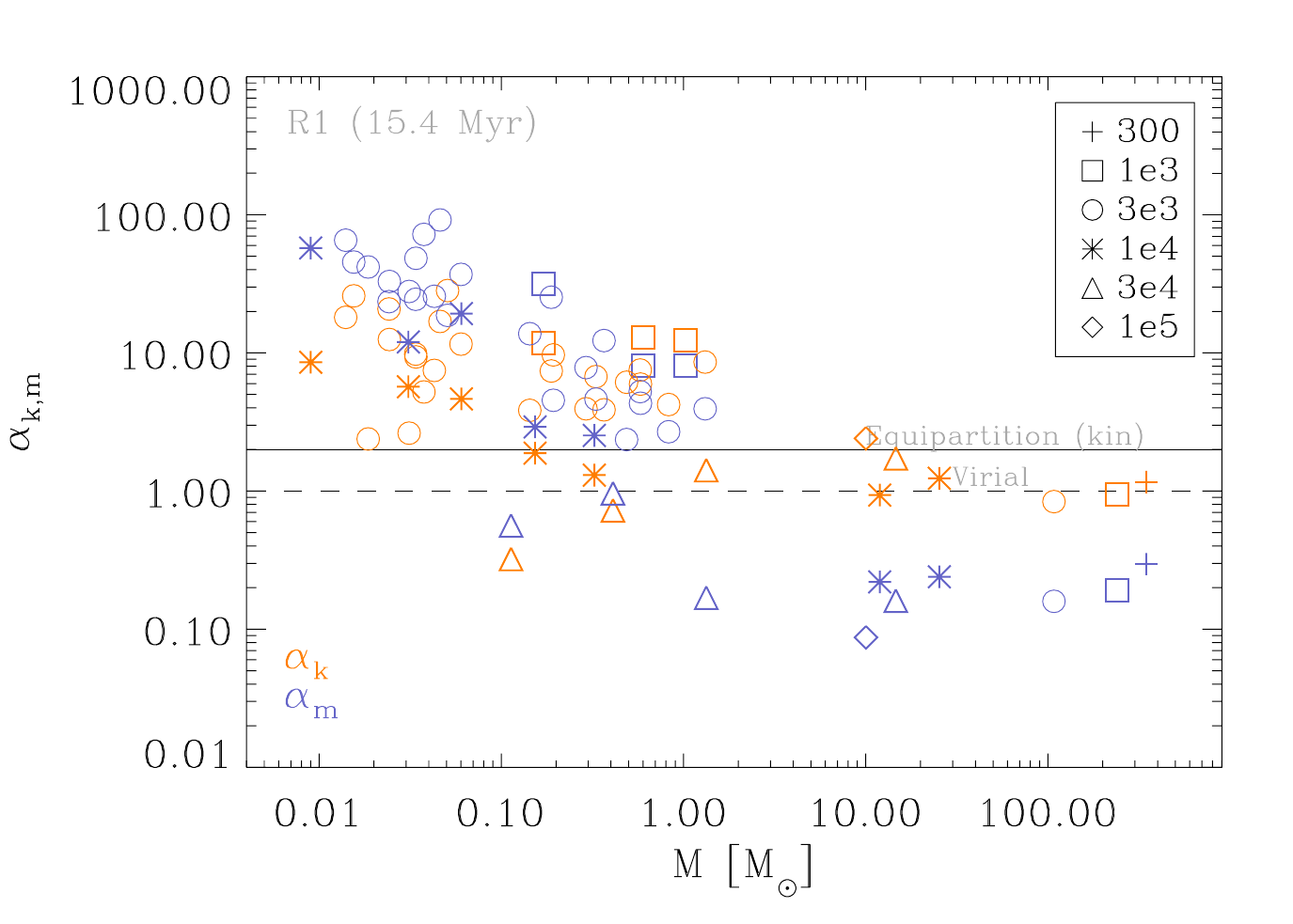}
    \caption{The evolution of $\Lcal$-$\Sigma$ (upper panels) and $\alpha$-$M$ (bottom panels) relation for region R1 at t=14.4 Myr (left panels, also shown in Figs.~\ref{fig:alphas_reg} and \ref{fig:Lratios_reg})  and t=15.4 Myr (right panels). At  the evolved time, the distribution of the clumps within R1 in these diagrams is similar to that of regions R2-R4 at t=14.4 Myr in Figs. \ref{fig:alphas_reg} and \ref{fig:Lratios_reg}, showing that the different subregions in the filament constitute an evolutionary sequence.}
   \label{fig:R1ev}
\end{figure*}

Evidence of the ongoing contraction of region R1 is presented in Figure \ref{fig:R1ev}, where we show its $\Lcal$-$\Sigma$ and $\alpha$-$M$ diagrams at time $t=15.4$ Myr (same as that of bottom-left panel in Fig.\ \ref{fig:sim}). At this time, R1 is seen to have similar distributions in these diagrams compared to the more evolved regions at the original time t=14.4 Myr. This is consistent with the evolutionary picture presented in Paper II.
Furthermore, R2 feeds R4, and R4 and R3 both feed R1, already from the beginning. At the same time, R1 grows into a hub denser than R4, and while regions R2 and R3 (as well as R4) continue accreting material from the surrounding medium, they are being driven through the filament towards R1.

On the other hand, at $t = 14.4$ Myr the more evolved regions (R3 and R4) of our hub-filament system  are seen, Fig.\ \ref{fig:Lratios_reg}, to contain a population of nearly-virial high-density clumps, both in the kinetic and the magnetic budget parameters. These clumps have already become locally self-gravitating by this time, and are therefore contracting due to their own self-gravity.

It is important to remark that, although the virial parameter and Larson ratio are seen to reach nearly virial values, this does not imply by any means that they are in virial equilibrium, since they are observed to be developing gravitational collapse centers in their interiors (although their sizes do not shrink, because at the same time they are accreting fresh material from their environment). Therefore, they should rather be described simply as reaching near energy equipartition {\it during} the compression-collapse process. 
 
From the above description, we can see that the individual regions are evolving, increasing their mass and the gravitational binding of their fragments, so that the latter gradually populate concave trajectories in the $\Lcal$-$\Sigma$ diagram. In addition, each one of them is found at a somewhat different evolutionary stage. Simultaneously, however, the various regions are part of the full-scale hub-filament system, which itself exhibits similar energy budget scalings, and in which the filaments are feeding the central hub. This illustrates the mechanism of \textit{collapses within collapses} proposed by the GHC scenario \citep{VS+19}.

It is important to note that this type of evolution, consisting of multi-scale gravitational contraction and a simultaneous increase in the SFR \citep[e.g.,] [] {Camacho+20} is central to the GHC scenario, which is represented by our simulation. In it, the turbulence is stirred self-consistently by various instabilities in the dense layer assembled by the converging flows \citep[e.g.,] [] {Walder+00, KI02, Audit+05, VS+06}, but it is insufficient to support the cloud, which therefore begins to undergo GHC soon after it surpasses its Jeans mass \citep[e.g.,] [] {VS+07}. Instead, in the gravoturbulent (GT) scenario, in which structures larger than dense molecular cores, of sizes $\gtrsim 0.1$ pc, are assumed to be supported by turbulence, stirred from external sources \citep[e.g.,] [] {MacLow+04, Padoan+16}, the clouds are stationary or even transient, rather than evolutionary \citep[e.g.,] [] {VS+00, MacLow+04, Padoan+20}, and are thus not expected to undergo any secular evolution. In the intermediate scenario proposed by \cite{Murray+15}, an increasing SFR is predicted but without cloud-scale collapse, and therefore, with no growth nor convergence of the intermediate-size hubs.

Moreover, the gravitationally-driven large-scale flow, exemplified in our simulation by the spatial approach between all four regions,\footnote{Note, however, that this approach is very slow. For example, a by-eye measurement of the motion of regions R1 and R3 shows that they approach each other at a speed of $\sim 0.6 \kms$, and moreover they do so while both moving together toward the left of the full region shown in Fig.\ \ref{fig:sim}, probably all participating in the infall toward the main gravitational potential 
trough at the center of the simulation. This is best illustrated by the animation accompanying this figure in the electronic edition.} is an essential feature of the GHC scenario, and is not expected to occur in scenarios where the large-scale flows are random, driven exclusively by supernova explosions, since in this picture, no large-scale collapse occurs.\footnote{\citet{Padoan+20} have suggested that large-scale flows of the hub-filament type can occur by pure inertial motions driven by supernova feedback, as illustrated in the sketch of their Fig.\ 1. However, this suggestion has not been backed up by numerical tests showing that such flows can indeed arise in the absence of self-gravity.} Finally, note that the approach between these regions, which eventually leads to the merging of some of them, can be viewed as the development of a cloud-cloud collision event \citep[see, for example, the review by] [] {Fukui+21}, albeit driven by large-scale gravity.

\subsection{The equivalence of temporal evolution and depth in the cores} \label{sec:equiv_time_depth}

In addition to the variation of the distribution of budget parameter across regions of different ages discussed in Sec.\ \ref{sec:scaling_age}, in Sec.\ \ref{sec:scaling_depth} we noted that, in the more evolved regions R3 and R4, hierarchically-nested clumps (i.e., clumps that are all within the same parent cloud, each defined by a different density threshold, such as the clumps indicated by the sequences of symbols from diamonds to pluses at high masses in the bottom panels of Fig.\ \ref{fig:alphas_reg}) have near equipartition values of both the kinetic and magnetic  virial parameters. Furthermore, $\akin$ for these clumps describes a moderately concave curve, as indicated by the lines connecting these symbols. 

Analogously, the panels for regions R3 and R4 in Fig.\ \ref{fig:Lratios_reg} show that the same sequences of nested clumps have nearly virial values of $\Lcalv$ and $\Lcal_{\rm m,rms}$ ($\Lcal_{\rm m,avg}$ is significantly lower and noisier), with again the kinetic parameter exhibiting a slightly concave curve consistent with the theoretical prediction of \citet[] [see their equation (8) and Fig.\ 2]{VS+19}, and with the trend observed for $\akin$ in Fig.\ \ref{fig:alphas_reg}.

The above results thus suggest an equivalence between the temporal evolution observed from panel to panel in Figs.\ \ref{fig:alphas_reg} and \ref{fig:Lratios_reg} (as well as the tracking of individual clumps over time, Paper II), and the radial stratification within a single clump, observed when thresholding it at different densities. This can be interpreted as a manifestation of the similarity property of gravitational collapse, in which the independent similarity variable, $\xi \equiv r/\cs t$, implies the equivalence of radial and temporal coordinate pairs that keep $\xi$ constant \citep[e.g.,] [] {WS85}. In a sense, material at higher density, deeper into a structure, is at a more evolved stage along the collapse flow than material further out, at lower densities.

\subsection{Gravitational driving of the kinetic and magnetic energies} \label{sec:grav_driv_kin_mag}

The ubiquitous approximate equipartition between the kinetic and gravitational energies found here strongly suggests that {\it{the velocity field in the clumps obtains its energy mainly from the gravitational contraction}}, since it appears virtually impossible that random, externally driven turbulence would manage to reach equipartition in every structure within the clouds. Of course, some truly random turbulence must be generated during the collapse \citep[e.g.,] [] {VS+98, Klessen+10, Robertson_Goldreich12}, but recent simulations show that the truly turbulent kinetic energy is a minority of the total kinetic energy, with the majority being in the infall motions \citep[e.g.,] [] {Guerrero+20}.

Furthermore, our finding that the magnetic budget parameters mimic closely the behavior of their kinetic counterparts suggests that {\it the magnetic field also obtains its energy from the collapse}. However, we have observed that the magnetic energy scales slightly more slowly with the gravitational energy ($\Em \sim \Eg^{0.65}$) than the kinetic one ($\Ek \sim \Eg^{0.85}$; see Fig.\ \ref{fig:energies}). If both the velocity and the magnetic field obtain their energy from the gravitational contraction, this implies that the transfer to magnetic modes is less efficient than to the kinetic ones. The origin of this disparity remains to be investigated, although we speculate that a possible reason is that the energy transfer to the magnetic modes occurs {\it via} the kinetic ones (the turbulence generates magnetic fluctuations) given that this mechanism cannot be 100\% efficient because some of the kinetic energy is lost to viscous (or, in our case, numerical) dissipation. 
Another possibility is that part of the magnetic energy can be generated, or amplified, by the compression of the magnetic field lines.\footnote{We thank the referee for this suggestion.} With this mechanism, the inefficiency would arise from the fact that compressions of the magnetic field lines do not occur from motions along them but rather from motions perpendicular to them. Actually, this possibility implies that the velocity field is not completely parallel to B.

\subsection{Comparison with previous work} \label{sec:compar_prev}

Although the scaling of the Larson ratio $\Lcal$ with column density $\Sigma$ was already recognized as a general trend for Galactic clouds several decades ago \citep{Keto+86}, in the last two decades numerical and observational works have identified this scaling as a generalization of Larson's relations \citep{Larson81}, not only for large Galactic clouds, but also for clouds in nearby galaxies, and for Galactic dense cores and clumps \citep[][]{Heyer+09, BP+11, BP+18, Leroy+15, Camacho+16, Camacho+20, Ibanez+16, Miville+17, Traficante+18a}. These works have explored physical properties considering two of the main energies, kinetic and gravitational. In the present work, we have presented a similar analysis considering the magnetic energy as well, extending the early results of \citet{Myers88a} and \citet{Mouschovias+95}, 
and of more recent, numerical \citep[e.g.,][]{Ibanez+22, Ganguly+22} 
and observational \citep{Liu+22} studies, which have found that the magnetic field's importance decreases with increasing gas density.

Figure \ref{fig:energies} shows that, similarly to the kinetic energy, the magnetic energy also increases with the gravitational energy, albeit with a slightly shallower slope (0.66 {\it vs.} 0.86). Correspondingly, the magnetic virial parameter also decreases with increasing clump mass, again similarly to the kinetic parameter, but with a steeper slope, as seen in Fig.\ \ref{fig:alphas}. Finally, the Larson ratio also exhibits a qualitatively similar trend with column density as the kinetic one, but again with the high-$\Sigma$ objects having a slight deficit of magnetic energy.
Another way to interpret the Larson ratio (Eq. \ref{eq:lrat}) is as a measure of the balance between gravitational and kinetic energies and, given that $\Lcal \propto \sqrt{\Sigma}$, 
$\Lcal$ can be used as an estimation of $\Eg$. Thus, since the relation between $\Lcal$ and B in \citet[Fig. 2]{Mouschovias+95} represents a measure of the magnetic field against gravity, this work shows equipartition between $\Em$ and $\Eg$. Assuming that B is proportional to $\Sigma$ (through the magnetic flux ratio), the Fig. 2 in \citet{Mouschovias+95} is equivalent to our $\Lcal$-plots, where we can observe in the numerical and observational data, that in some cases both samples satisfy equipartition. Although in our case, we interpret the equipartition as a consequence of the gravitational collapse during the clump's evolution.

The decrease of the kinetic virial parameter with increasing clump mass is well documented observationally \citep[e.g.,] [] {Kauffmann+13, Miville+17, Traficante+18a, Liu+22}, as well as in non-magnetic simulations at the few hundred parsec scale \citep[e.g.,] [] {Camacho+16}, and magnetic simulations at the kpc scale \citep[e.g.,] [] {Ibanez+16}. Here we observe that the {\it magnetic} virial parameter exhibits the same trend. 

Moreover, because the magnetic virial parameter is the inverse of the squared mass-to-flux ratio (c.f.\ eq.\ [\ref{eq:fmr}]), this result implies that the mass-to-flux ratio increases with increasing mass, in particular for objects nested hierarchically. This effect was first detected observationally by \citet{Crutcher+09}, and derived analytically by \citet{Gomez+21} as a general feature of cores with a certain density profile and a magnetic field strength that increases with increasing density. Naively, this trend could be interpreted as implying that the inner, denser structures are more strongly magnetically supported, while in reality it only means that they are being crushed by the infall of the outer parts of a clump onto them. This is confirmed by the fact that, the low column density clumps in region R1 (considering $B_\mathrm{rms}$, indicated by the blue symbols) appear magnetically subcritical in Fig.\ \ref{fig:bsigmas_reg}, yet they are all part of the same structure, which is  supercritical at thresholds $\nth = 300$, 1000, and 3000 $\pcc$ (plus, square and circle symbols, respectively).
This is also consistent with slightly subcritical mass-to-flux ratio measurements found in massive dense cores with active star formation \citep[e.g.,][]{Juarez+17,Soam+19,Devaraj+21,Ngoc+21,Palau+21,Hoang+22}.

\subsection{Comparison with observations} \label{sec:obs}

The results obtained so far can be directly compared to observational works. In particular, \citet{Palau+21} studied the magnetic field properties in a sample of 18 Galactic massive dense cores, and analyzed them in a uniform way. The cores in this sample present star formation activity in a very early evolutionary stage, where stellar feedback (either from outflows or from ionizing photons) is not strongly affecting the dynamics yet. Therefore, the \citet{Palau+21} sample is well-suited to be compared to the simulations presented here.

In order to calculate the magnetic field strength of the cores of the sample, we used here the recently published Skalidis-Tassis (ST) method \citep{SkalidisTassis21, Skalidis+21b} instead of the traditional Davis-Chandrasekhar-Fermi (DCF) method \citep{Davis51, Chandra53}. Since the ST method takes into account the compressible modes of turbulence, while the DCF method assumes that only the incompressible MHD waves (Alfv\'en waves) produce the dispersion in polarization position angles, the ST method should provide a more accurate estimate of the magnetic field strength \citep{Skalidis+21b}. The ST method, similarly to the DCF method, requires the knowledge of the density $\rho$, velocity dispersion of the turbulent motions $\sigma_\mathrm{turb}$, and polarization position angle dispersion $\sigma_\mathrm{PA}$ of the core. The magnetic field strength then scales as the inverse of the square root of the polarization angle dispersion (while in the DCF method the magnetic field strength scales directly as the inverse of the polarization angle dispersion):

\begin{equation}
  B_\mathrm{0} \approx \sqrt{2\pi\rho}\, \frac{\sigma_\mathrm{turb}}{\sqrt{\tan \sigma_\mathrm{PA}}}.
  \label{eq:ST}
\end{equation}

In this equation, $\tan \sigma_\mathrm{PA}$ was used instead of $\sigma_\mathrm{PA}$, to take into account that in some cases $\sigma_\mathrm{PA}$ is large \citep[$\gtrsim 25^\circ$,][]{FalcetaGoncalves+08}{}{}.
We applied equation~\eqref{eq:ST} using the results from \citet{Palau+21} to obtain the  magnetic field strength, $\Ek$, $\Em$, $\Eg$, and the virial parameters, listed in the Appendix along with further details of the calculations.

The resulting observational parameters are overplotted in Figs.~\ref{fig:energies}--\ref{fig:bfield}, indicated by filled circles with error bars, on the top of the simulated cores. In general, the cores of \citet{Palau+21} lie in the high $\Eg$ range, as expected from the fact that these cores are already undergoing star formation. Figure~\ref{fig:energies} shows that, for the observed cores, $\Ek$ follows a one-to-one relation with $\Eg$ as in the simulations, and $\Em$ on average also follows the shallower trend of the simulated cores. Similarly, Figs.~\ref{fig:alphas}-\ref{fig:bfield} reveal in general a good consistency between the observed and simulated data.
Moreover, the simulations presented here were compared with a very recent compilation of magnetic field measurements in low- and high-mass dense clumps and cores \citep{Liu+22}, which we complemented with the very recent work of \citet{Chung+22}. Although such a comparison is worth it because it includes a very large number of observed clumps/cores, the fact that several methods were applied to different objects of the sample makes the overall comparison more risky. Even having such an inhomogeneous sample, still good consistency between the observed and simulated virial parameters is found (see the Appendix for further details).

Finally, the main result of \citet{Palau+21} that the magnetic field seems to suppress fragmentation in the observed massive dense cores was explored in the simulations. In our simulated data at 14.4 Myr, the region R4 presents a higher fragmentation level having already formed 2 sinks, while the region R1 presents no sinks yet. Thus, it would be expected that the average mass-to-flux ratio of R4 was larger than the mass-to-flux ratio of R1. This is precisely what we measure, a mass-to-flux ratio $\mu_{\rm avg} = 6.2$ (using the mean field) for R4 ($\mu_{\rm rms} = 2.0$ using the rms value of the field) and $\mu_{\rm avg} = 3.2$ ($\mu_{\rm rms} = 1.5$) for R1, in agreement with observations and a wealth of previous theoretical and a numerical work \citep[see the introduction of][for a list of references]{Palau+21}.

Note, however, that this does not necessarily imply that the magnetic field has any direct impact on the fragmentation level, as it is possible that accretion onto the core has the dual effect of increasing its mass,\footnote{Previous numerical simulations \citep[Paper II; ] [] {Gonzalez+20} have shown that cores defined by some fixed boundary (either a density or intensity threshold, or a fixed Eulerian boundary, such as a beam size) continue to increase their mass by accretion even after beginning to form stars, until feedback destroys the filamentary accretion flow.} thus increasing the number of Jeans masses it contains as well as its mass-to-flux ratio. In this sense, both the fragmentation level and the mass-to-flux ratio can be considered to be controlled by the gravitational accretion flow starting at the largest scale.

\subsection{Caveats and convergence} \label{sec:caveats}

Our results are of course dependent on a number of procedural choices and limitations, which we enumerate now. First, as mentioned in Sec.\ \ref{sec:sample}, we do not use a branching-based clump finding algorithm such as {\sc Dendrograms}, because we want the density levels of the clumps to be consistent among the different regions we investigate. In contrast, in many observational studies, for example, cores are defined as the leaves (the highest branching stage) of a hierarchically nested sequence of objects defined through the {\sc Dendrograms} algorithm. This means that the density (or column density) threshold is not necessarily consistent for cores appearing in different regions of the hub-filament system. Our choice avoids this source of randomness, but on the other hand requires that comparisons with observations must be performed in terms of the physical properties of the identified objects, rather than by branching level.

A second limitation is that our simulation does not include feedback, which is the reason behind the selection criterion of clumps not containing too high a sink mass, so that our selected clumps should not be too strongly affected by the feedback expected from the sinks. 
This means that the results of this work must be compared to observational works focused on samples of clumps and cores in very early evolutionary stages.

\begin{figure}
	\includegraphics[width=\columnwidth]{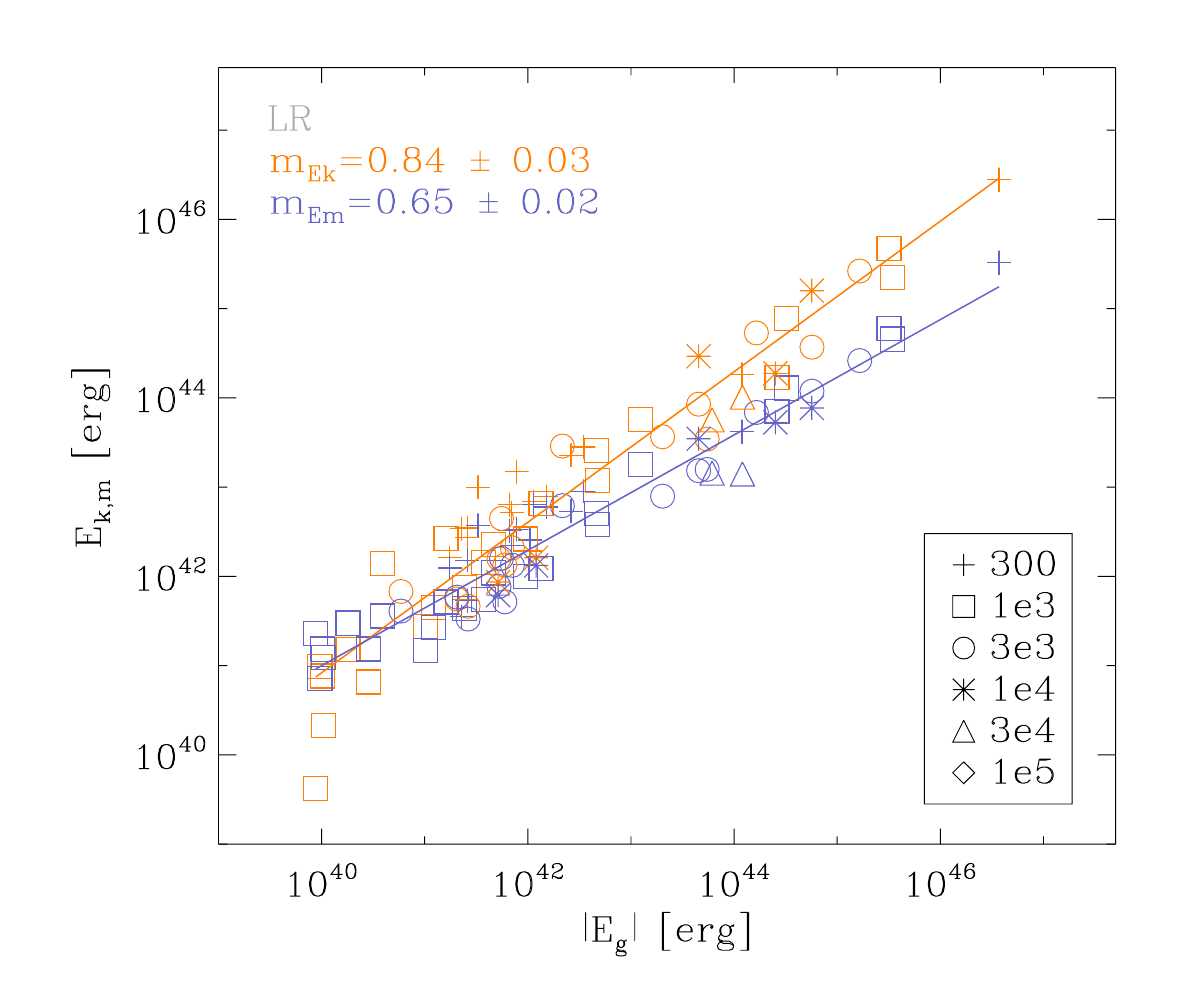}
    \caption{Scaling of the kinetic (orange symbols) and magnetic energies (blue symbols), in a simulation identical to the one we analyzed, but with four times lower resolution  (0.03 pc).}
   \label{fig:energies_LR}
\end{figure}

Third, our result that the magnetic energy scales with gravitational energy with a shallower slope than the kinetic energy (Fig.\ \ref{fig:energies}) could easily be originated by a different numerical diffusion rate for the velocity and the magnetic fields. Therefore, it is important to perform a convergence test, to determine whether the scaling slopes for the two energies depend on resolution. For economy, we check for convergence using a simulation with {\it lower} resolution. In Fig.\ \ref{fig:energies_LR} we show the scaling of both energies in the original simulation by \citet{Zamora+18}, which has a four times lower resolution. We see that, within the uncertainties, the slopes are the same as those determined at high resolution (Fig.\ \ref{fig:energies}), strongly suggesting that the different scaling of the kinetic and magnetic energies with the gravitational energy is a real, physical effect, and not due to numerical diffusion.

Finally, we note that the gravitational energies we have computed in this work, given by eq.\ \eqref{eq:Eg}, assume the prescription for a uniform-density sphere, and therefore actually constitute a {\it lower limit} to the actual self-gravitating energy of the clumps, as objects that are centrally concentrated have larger values of $\Eg$ than uniform objects of the same mass and radius \citep{BP+18}.

\section{Summary and conclusions}
\label{sec:concl}

In this paper, we have investigated the kinetic and magnetic energy budget in molecular clouds undergoing gravitational fragmentation in a numerical simulation. We studied a hub-filament system arising self-consistently during the gravitational contraction and fragmentation of a giant molecular\footnote{The simulation does not track the chemistry in the cloud, so we only label it as molecular given its physical properties, like mass, temperature, and mean density.} cloud, which in turn formed in a  converging-flow large-scale simulation with initial conditions corresponding to the lukewarm atomic medium. We have investigated the properties of the clumps (generically interpreted as density enhancements, regardless of their size scale) in four sub-regions of the system, constituting a hierarchy of hub-filament systems.
Indeed, two hubs are identified at different epochs. The first one, region R4, at t=14.4 Myr has the highest-density fragments and the highest mass in sink particles. Later, at t=15.4 Myr, region R1, fed by the gravitationally-driven gas flow along the filament begins its star formation activity becoming the dominant center of collapse. 

To investigate the energy budget, we have considered the standard kinetic Larson ratio, $\Lcalv$, and virial parameter, $\akin$, respectively defined by eqs.\ \eqref{eq:lrat} and \eqref{eq:ak}, as well as their magnetic counterparts, $\Lcalb$ and $\amag$, respectively defined by eqs.\ \eqref{eq:lmag} and \eqref{eq:fmr}. We refer to $\alpha$ and $\Lcal$ collectively as the {\it energy budget parameters}. We also investigated the scaling of the kinetic and magnetic energies with the gravitational one.

Our main results are as follows:

\begin{itemize}

\item The sequence of regions R1-R4 constitutes a sequence of increasing star formation rate, which we show correspond to different evolutionary stages.

\item The four regions simultaneously evolve by themselves and as a part of the whole system, in agreement with the {\it collapses within collapses} concept in the GHC scenario. Specifically, each region increases its mass and star formation activity, acting as local hubs, while simultaneously regions R2 and R3 flow along larger-scale filaments that feed regions R1 and R4. This large-scale, gravitationally-driven accretion flow is an essential feature of the GHC scenario.

\item The kinetic and magnetic energies of clumps in all four regions are close to equipartition with the gravitational energy, although with a slight excess in objects with the lowest gravitational energies. This is consistent with objects at all scales starting out supervirial as a consequence of an external compression, and then becoming locally gravitationally bound. At this point, the energy released during the gravitational contraction is distributed evenly between the kinetic and magnetic modes.

\item For hierarchically nested objects, the larger parent structures appear more strongly bound than their daughter substructures, indicating that the external compressions that form the latter are due to the gravitational contraction of the former.

\item The magnetic $\alpha$ and $\Lcal$ parameters exhibit similar scalings with column density and mass as their kinetic counterparts, although the ratio $\Em/\Ek$ decreases for increasing $|\Eg|$. This suggests that both the velocity and the magnetic fields derive their energy from the gravitational contraction, although the energy transfer to the magnetic field is slightly less efficient than that to the velocity field.

\item The magnetic Larson ratio calculated using the mean value of the magnetic field within the clumps is often significantly lower than that using the rms value of the field. This implies that a significant, and often dominant part of the magnetic energy is in the field fluctuations within the clumps rather than in the mean field.

\item In agreement with our previous results (Paper I and Paper II) we find that the evolution of the individual regions gradually populate the \LS\ and \am\ diagrams toward higher column densities and masses, respectively, in agreement with the prediction from the GHC scenario \citep{VS+19}.

\item The observational data of massive clumps continue the trends of our numerical clumps in the \LS\ and \am\ diagrams toward larger column densities and masses, respectively. This is probably due to the fact that we have considered relatively early evolutionary stages of the simulation, in which the analogs of massive clumps have not formed yet.

\end{itemize}

We conclude that the consistency between the energy budgets in our numerical clump sample (extracted from a simulation undergoing global hierarchical collapse), and in the various observational clump samples we have considered, supports GHC as the controlling mechanism in real molecular clouds, and the possibility that the velocity and magnetic fields in the cloud substructures are driven mainly by gravity, rather than opposing it. \\

\section*{Acknowledgements}

We thank the referee, Mordecai Mac Low, for insightful comments and suggestions that greatly improved the paper, and useful comments from Javier Ballesteros-Paredes and Junhao Liu.
V.C. and M.Z.A. acknowledge support from CONACyT grant number A1-S-54450 to Abraham Luna Castellanos (INAOE) and from CONACYT grant number 2409949.
A.P. and E.V.-S. acknowledge financial support from UNAM-PAPIIT grant IG100223. 
A.P. further acknowledges support from Sistema Nacional de Investigadores of CONACyT and UNAM-PAPIIT IN111421 grant. 
A.P. and M.Z.A. acknowledge financial support from CONACyT grant number 86372 within the ``Ciencia de Frontera 2019'' program, entitled `Citlalc\'oatl: A multiscale study at the new frontier of the formation and early evolution of stars and planetary systems’, M\'exico.
Finally, M.Z.A. acknowledge financial support from CONACyT grant number 320772 of the call ``Ciencia Básica y/o Ciencia de Frontera. Modalidad: Paradigmas y Controversias de la Ciencia 2022'.
The numerical simulations were performed in the Laboratorio Nacional de Supercómputo del Sureste de México (LNS; a member of the CONACYT network of national laboratories) and the Miztli cluster at DGTIC-UNAM through proposal LANCAD-UNAM-DGTIC-188. 
The visualization was carried out with the {\tt yt} software \citep{yt}.

\section*{Data Availability}

The data generated for this article will be shared on reasonable request to the corresponding author.



\bibliographystyle{mnras}



\appendix

\section{Parameters of the observational sample}

\citet{Palau+21} study the polarization properties from thermal dust emission of a sample of massive dense cores undergoing intermediate/high-mass star formation. The original sample included 18 cores, but for two of them the polarized emission was too faint to infer any physical quantity, and the final sample with inferred magnetic field strength contains 16 cores.

To estimate the magnetic field strength from equation~\ref{eq:ST} for our observational data  \citep{Palau+21}, we took the average density reported in column (9) of Table~3 of \citet{Palau+21}, the turbulent velocity dispersion calculated as $\sigma_\mathrm{turb}=Q\,\sigma_\mathrm{nonth}$ (with $Q\sim0.5$ and $\sigma_\mathrm{nonth}$ being the non-thermal velocity dispersion taken from column (4) of Table~4 of Palau et al. 2021), and $\sigma_\mathrm{PA}$ taken from column (2) of Table~5 of \citet{Palau+21}, which is estimated by calculating the standard deviation of the weighted mean of the polarization position angles. We list the values used to obtain the magnetic field strength in Table A1, along with the sonic Mach number and the Alfvénic Mach number. In Table A2 we report the mass-to-flux ratio, the kinetic, magnetic, and gravitational energies, and the kinetic and virial parameters. The resulting plots using these data are shown in Figs.~\ref{fig:energies}--\ref{fig:bfield}.

\begin{table*}
\centering
\caption{Mass, surface density and parameters of the Palau et al. sample used to infer the magnetic field strength \label{tab:P21sampleB}}
\setlength{\tabcolsep}{4pt}
\begin{tabular}{lccccccccccccccc}
\hline
&\multicolumn{1}{c}{$T_\mathrm{0.15pc}$$^\mathrm{a}$}
&\multicolumn{1}{c}{$M_\mathrm{0.15pc}$$^\mathrm{a}$}
&\multicolumn{1}{c}{$n_\mathrm{0.15pc}$$^\mathrm{a}$}
&\multicolumn{1}{c}{$\Sigma_\mathrm{0.15pc}$$^\mathrm{a}$}
&\multicolumn{1}{c}{$\sigma_\mathrm{nonth}$$^\mathrm{a}$}
&\multicolumn{1}{c}{$\sigma_\mathrm{PA}$$^\mathrm{a}$}
&\multicolumn{1}{c}{$B_\mathrm{ST21}$$^\mathrm{a}$}
\\
Source
&\multicolumn{1}{c}{(K)}
&\multicolumn{1}{c}{($\Msun$)}
&\multicolumn{1}{c}{($10^5$\,cm$^{-3}$)}
&\multicolumn{1}{c}{($10^{22}$\,cm$^{-2}$)}
&\multicolumn{1}{c}{(km\,s$^{-1}$)}
&\multicolumn{1}{c}{($^\circ$)}
&\multicolumn{1}{c}{(mG)}
&\multicolumn{1}{c}{$\mathcal{M}_\mathrm{s}$$^\mathrm{a}$}
&\multicolumn{1}{c}{$\mathcal{M}_\mathrm{A}$$^\mathrm{a}$}
\\
\hline
W3IRS5  &$118\pm14$ &$22\pm3$   &$1.8\pm0.2$    &$5.6\pm0.7$    &$1.35\pm0.14$  &$50\pm7$   &$0.14\pm0.02$  &$3.59$ &$5.34$ \\
W3H2O   &$82\pm9$   &$59\pm8$   &$4.8\pm0.6$    &$15\pm2$       & $2.8\pm0.3$   &$41\pm11$  &$0.57\pm0.13$  &$9.06$ &$4.57$ \\
G192    &$42\pm4$   &$11\pm2$   &$0.9\pm0.1$    &$2.7\pm0.4$    & $1.03\pm0.10$ &$18\pm6$   &$0.15\pm0.03$ &$4.59 $ &$2.79$ \\
N6334V  &$56\pm6$   &$51\pm12$  &$4.2\pm1.0$    &$13\pm3$       & $2.1\pm0.2$   &$35\pm6$   &$0.43\pm0.08$ &$7.95$ &$4.14$ \\
N6334A  &$40\pm4$   &$46\pm7$   &$3.8\pm0.6$    &$12\pm2$       & $0.56\pm0.06$ &$16\pm6$   &$0.18\pm0.04$ &$2.56$ &$2.60$ \\
N6334I  &$70\pm8$   &$73\pm13$  &$6.0\pm1.0$    &$18\pm3$       & $0.97\pm0.10$ &$7\pm3$    &$0.60\pm0.17$ &$3.35$ &$1.65$ \\
N6334In &$52\pm6$   &$81\pm14$  &$6.6\pm1.0$    &$20\pm4$       & $1.39\pm0.14$ &$6\pm3$    &$1.0\pm0.3$ &$5.57$ & $1.59$ \\
G34-0   &$46\pm4$   &$44\pm9$   &$3.6\pm0.7$    &$11\pm2$       & $1.07\pm0.11$ &$17\pm6$   &$0.32\pm0.08$ &$4.56$  &$2.67$ \\
G34-1   &$33\pm4$   &$42\pm8$   &$3.4\pm0.6$    &$11\pm2$       & $0.84\pm0.08$ &$6\pm2$    &$0.41\pm0.09$ &$4.22$ &$1.61$ \\
G35     &$46\pm4$   &$36\pm5$   &$3.0\pm0.4$    &$9.1\pm1.2$    & $2.3\pm0.2$   &$33\pm5$   &$0.42\pm0.06$ &$9.63 $ &$3.93$ \\
CygX-N3 &$23\pm2$   &$29\pm4$   &$2.4\pm0.3$    &$7.4\pm0.9$    & $0.94\pm0.09$ &$47\pm16$  &$0.12\pm0.04$ &$5.66$ &$5.04$ \\
W75N    &$67\pm7$   &$48\pm7$   &$4.0\pm0.5$    &$12\pm2$       & $2.0\pm0.2$   &$35\pm16$  &$0.40\pm0.13$ &$7.03$ &$4.12$ \\
DR21OH  &$42\pm4$   &$110\pm20$ &$8.6\pm1.7$    &$26\pm5$       & $2.0\pm0.2$   &$32\pm13$  &$0.63\pm0.18$ &$8.83$ &$3.89$ \\
CygX-N48&$31\pm3$   &$56\pm11$  &$4.6\pm0.9$    &$14\pm3$       & $1.22\pm0.12$ &$43\pm12$  &$0.23\pm0.06$ &$6.33$ &$4.77$ \\
CygX-N63&$27\pm2$   &$20\pm3$   &$1.7\pm0.3$    &$5.1\pm0.9$    & $1.00\pm0.10$ &$46\pm20$  &$0.11\pm0.04$ &$5.56$ &$4.97$ \\
N7538S  &$49\pm3$   &$72\pm5$   &$5.9\pm0.5$    &$18.1\pm1.4$   & $2.6\pm0.3$   &$38\pm4$   &$0.61\pm0.08$ &$10.81$ &$4.34$ \\
\hline
\end{tabular}

$^\mathrm{a}$ All the properties are averaged within a field of view of 0.15 pc, which is the common field of view for all the regions, and are taken from Tables 3, 4, and 5 of \citet{Palau+21}, except for the magnetic field strength and the Alfvén Mach number. The non-thermal velocity dispersion $\sigma_\mathrm{nonth}$ is obtained from the line width measured by fitting a Gaussian to the HCO$^+$\,(4--3) spectrum averaged over a region of 0.15~pc of diameter \citep{Palau+21}, and a factor of $Q=0.5$ was applied to obtain $\sigma_\mathrm{turb}$, the dispersion used to estimate the magnetic field strength \citep[$\sigma_\mathrm{turb}=Q\sigma_\mathrm{nonth}$,][]{Palau+21}. The dispersion of the polarization position angles is obtained from the standard deviation of the weighted mean (column (2) of Table 5 of \citet{Palau+21}). The magnetic field strength $B_\mathrm{ST21}$ is obtained by applying the ST method (equation~\ref{eq:ST}) from \citet{SkalidisTassis21, Skalidis+21b}. The sonic Mach number, $\mathcal{M}_\mathrm{s}$, was calculated as $\mathcal{M}_\mathrm{s}=\sqrt{3}\sigma_\mathrm{nonth}/c_\mathrm{s}$, with $c_\mathrm{s}$ being the sound speed, $c_\mathrm{s}=\sqrt{k_\mathrm{B}\,T/(m_\mathrm{w}\,m_\mathrm{H})}$, $m_\mathrm{w}=2.3$ (the mean molecular weight) and $T$ taken from column 2 of this table \citep{Palau+21}. The Alfvénic Mach number, $\mathcal{M}_\mathrm{A}$, was estimated as $\mathcal{M}_\mathrm{A}=\sqrt{3}\sigma_\mathrm{nonth}/v_\mathrm{A}$, where the Alfvén velocity is $v_\mathrm{A}=\sqrt{B^2/4\pi\rho}$ (in cgs units), with the magnetic field $B$ taken from column 8 of this table, and the density $\rho$ taken from column 4 of this table and converted to mass density using $m_\mathrm{w}=2.8$ \citep{Palau+21}.

\end{table*}

\begin{table*}
\centering
\caption{Mass-to-flux ratios, energies and virial parameters of the Palau et al. sample \label{tab:P21sampleenergies}}
\setlength{\tabcolsep}{4pt}
\begin{tabular}{lcccccccccccc}
\hline
&&\multicolumn{1}{c}{$\Ek$$^\mathrm{a}$}
&\multicolumn{1}{c}{$\Em$$^\mathrm{a}$}
&\multicolumn{1}{c}{$\Eg$$^\mathrm{a}$}
\\
Source
&\multicolumn{1}{c}{$\mu$$^\mathrm{a}$}
&\multicolumn{1}{c}{($10^{45}$\,erg)}
&\multicolumn{1}{c}{($10^{45}$\,erg)}
&\multicolumn{1}{c}{($10^{45}$\,erg)}
&\multicolumn{1}{c}{$\alpha_\mathrm{k}$}
&\multicolumn{1}{c}{$\alpha_\mathrm{m}$}
\\
\hline
W3IRS5  &$2.8\pm0.6$   &$1.2\pm0.3$ &$0.04\pm0.02$    & $0.34\pm0.08$   &$7\pm2$      &$0.11\pm0.05$\\
W3H2O   &$1.9\pm0.5$   &$14\pm3$    &$0.7\pm0.3$      & $2.3\pm0.6$     &$12\pm3$     &$0.29\pm0.15$\\
G192    &$1.3\pm0.3$   &$0.3\pm0.1$ &$0.05\pm0.02$    & $0.07\pm0.02$   &$9\pm2$      &$0.6\pm0.3$\\
N6334V  &$2.2\pm0.7$   &$7\pm2$     &$0.38\pm0.14$    & $1.8\pm0.8$     &$7\pm2$      &$0.21\pm0.12$\\
N6334A  &$4.8\pm1.3$   &$0.4\pm0.1$ &$0.07\pm0.03$    & $1.5\pm0.4$     &$0.6\pm0.2$  &$0.04\pm0.02$\\
N6334I  &$2.2\pm0.7$   &$2.1\pm0.6$ &$0.8\pm0.4$      & $3.6\pm1.3$     &$1.2\pm0.3$  &$0.21\pm0.14$\\
N6334In &$1.6\pm0.5$   &$4.6\pm1.2$ &$1.9\pm1.1$      & $4.4\pm1.5$     &$2.1\pm0.6$  &$0.4\pm0.3$\\
G34-0   &$2.5\pm0.8$   &$1.5\pm0.4$ &$0.21\pm0.10$    & $1.3\pm0.5$     &$2.3\pm0.6$  &$0.16\pm0.10$\\
G34-1   &$1.9\pm0.6$   &$0.9\pm0.2$ &$0.34\pm0.15$    & $1.2\pm0.4$     &$1.5\pm0.4$  &$0.28\pm0.16$\\
G35     &$1.6\pm0.3$   &$5.5\pm1.3$ &$0.36\pm0.10$    & $0.9\pm0.2$     &$12\pm3$     &$0.40\pm0.15$\\
CygX-N3 &$4.4\pm1.4$   &$0.8\pm0.2$ &$0.03\pm0.02$    & $0.58\pm0.14$   &$2.6\pm0.6$  &$0.05\pm0.03$\\
W75N    &$2.2\pm0.7$   &$5.8\pm1.4$ &$0.3\pm0.2$      & $1.6\pm0.4$     &$7.2\pm1.8$  &$0.21\pm0.15$\\
DR21OH  &$3.0\pm1.1$   &$12\pm3$    &$0.8\pm0.5$      & $8\pm3$         &$3.3\pm0.9$  &$0.11\pm0.07$\\
CygX-N48&$4.4\pm1.4$   &$2.5\pm0.7$ &$0.11\pm0.06$    & $2.2\pm0.8$     &$2.3\pm0.6$  &$0.05\pm0.03$\\
CygX-N63&$3.4\pm1.4$   &$0.6\pm0.2$&$0.03\pm0.02$     & $0.27\pm0.09$   &$4.3\pm1.2$  &$0.09\pm0.07$\\
N7538S  &$2.1\pm0.3$   &$15\pm3$    &$0.8\pm0.2$      & $3.5\pm0.5$     &$8.4\pm1.8$  &$0.22\pm0.06$\\
\hline
\end{tabular}

$^\mathrm{a}$ $\mu$ is the mass-to-magnetic flux ratio and is calculated as $\mu=\Sigma\,m_\mathrm{w}\,m_\mathrm{H}/(B\,\lambda_\mathrm{crit})$, with $\mu_\mathrm{crit}=\sqrt{5/(18\pi^2\,G)}$, as given after equation~\ref{eq:fmr} in Sec.~\ref{sec:defs}, and $m_\mathrm{w}=2.8$. $\Sigma$ and $B$ are taken from Table~\ref{tab:P21sampleB}. $E_\mathrm{k}$ is the total kinetic energy calculated as $\Ek=\frac{3}{2}M_\mathrm{0.15pc}\sigma_\mathrm{nonth}^2$, $E_\mathrm{m}$ is the magnetic energy calculated as $\Em=\frac{B^2}{6}R^3$, and $\Eg=$ is the gravitational energy calculated as $\Eg=\frac{3}{5}G\,M_\mathrm{0.15pc}^2/R$, with $R=0.15/2=0.075$~pc, and $M_\mathrm{0.15pc}$, $\sigma_\mathrm{nonth}$, and $B$ taken from Table~\ref{tab:P21sampleB}. 

\end{table*}

\begin{figure*}
	
    \includegraphics[width=\columnwidth]{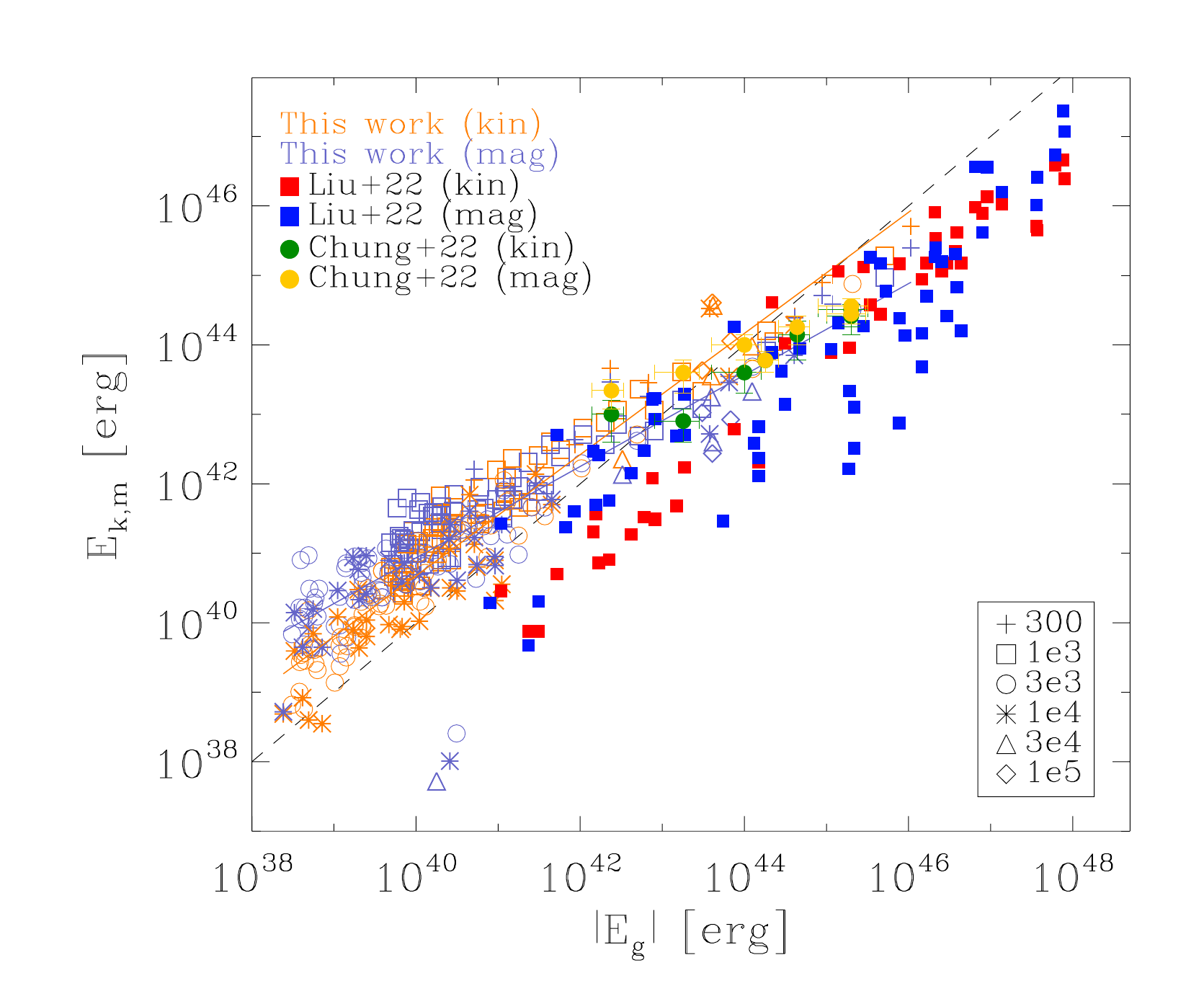}
    \includegraphics[width=\columnwidth]{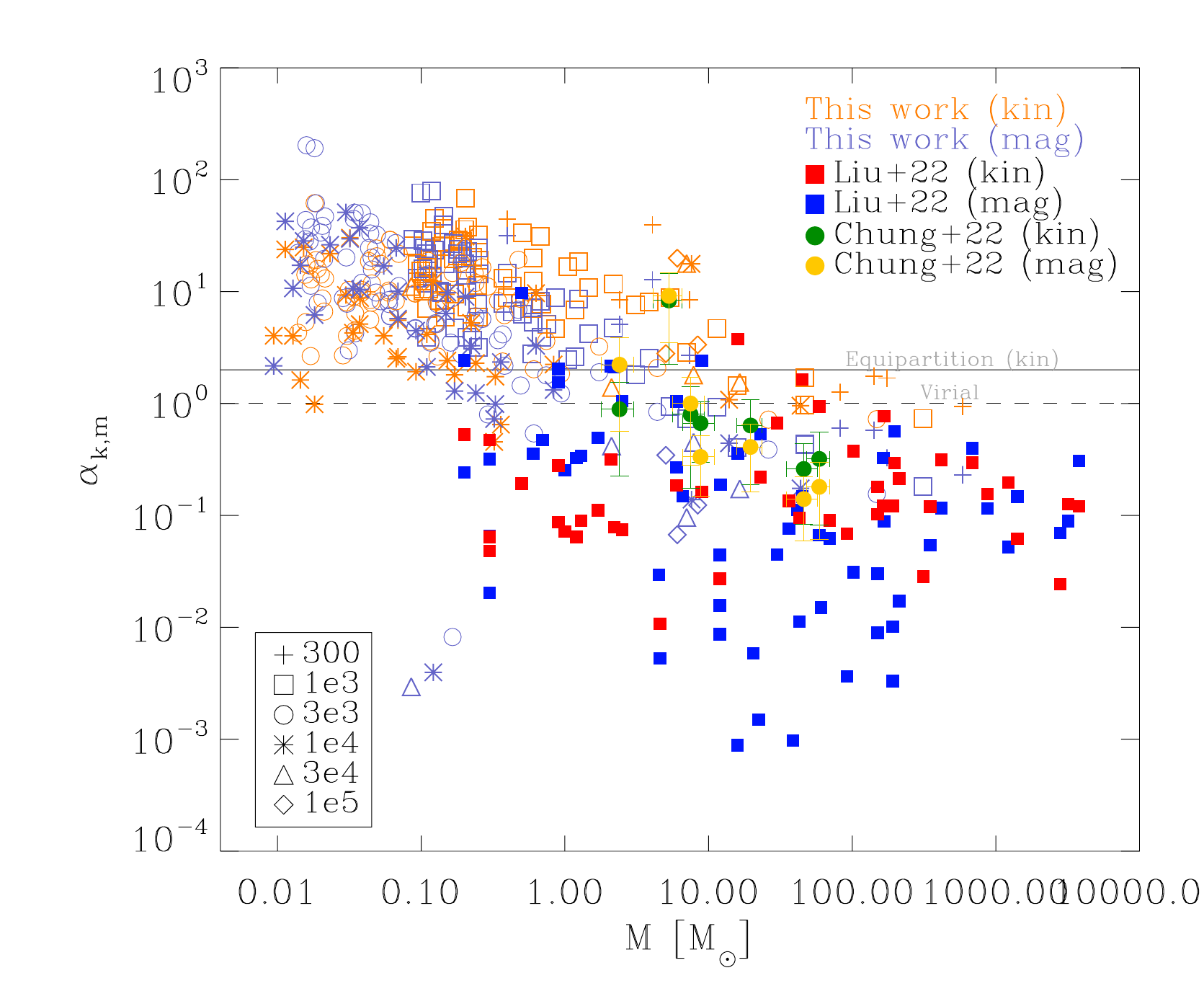}
    
    \caption{ Comparison between observations and the simulation in this work. The filled square symbols correspond to the data presented in \citet{Liu+22}, for which energies (left panel, $\Eg, \Ek,$ and $\Em$) were computed following the equations in table \ref{tab:P21sampleenergies}, while filled circles represent the data of the filament-hub system in \citet{Chung+22} with their corresponding errors, taken from their Table 2. In both samples red symbols represent the kinetic case and the blue symbols the magnetic one. The right panel shows the virial parameter computed according to our Eqs.(\ref{eq:ak}) and (\ref{eq:am}) for both samples.}
   \label{fig:alphasamples}
\end{figure*}

We additionally compare the results on the \am\ scaling in our numerical data with the observational data collected in the review by \citet{Liu+22}, which were complemented with the recent data of \citet{Chung+22} (Fig. \ref{fig:alphasamples}). In the former, the magnetic field is derived using the Davis-Chandrasekhar-Fermi method.
In the latter reference, the authors study the fragmentation of a hub-filament system by computing their kinetic, gravitational, and magnetic energies \citep[see Sec. 4.2][]{Chung+22}. 
For comparison between this work and the observational data, we computed the energies and the virial parameters according to the procedure in this work. Using the data provided in Table 1 of \citet{Liu+22}, energies were derived according to the equations in our Table \ref{tab:P21sampleenergies}, while energies for \citet{Chung+22}, were taken directly from their Table 2 with their corresponding errors. The kinetic and magnetic virial parameters for both samples were computed according to equations \ref{eq:ak} and \ref{eq:am} of this work, respectively.
 Although the observed samples of \citet{Liu+22,Chung+22} are not homogeneous, a reasonable consistency between the observed and the simulated clumps and cores can be appreciated in Fig.~\ref{fig:alphasamples}.

\bsp	
\label{lastpage}
\end{document}